\documentclass[preprint,12pt]{elsarticle}
\usepackage{graphics}
\usepackage{graphicx}
\usepackage{epsfig}
\usepackage{amssymb}
\journal{Elsevier Science}
\begin{document}

\begin{frontmatter}
\title{Aggregation and Probabilistic Verification for Data Authentication in VANETs\tnoteref{label1}}
\tnotetext[label1]{Research supported by the Spanish MINECO and the European FEDER Fund under Projects TIN2011- 25452 and IPT-2012-0585-370000; and the FPI scholarships BES-2009-016774 and ACIISI-BOC Number 60.}
\author{J. Molina-Gil\corref{cor1}}
\ead{jmmolina@ull.es}
\author{P. Caballero-Gil\corref{cor2}}
\ead{pcaballe@ull.es}
\author{C. Caballero-Gil\corref{cor3}}
\ead{ccabgil@ull.es}
\cortext[cor2]{Corresponding author. Tel: +034 922 318 176 }
\address{Department of Statistics, Operations Research and Computing,
\\ University of La Laguna,
\\ 38271 Tenerife, Spain\fnref{label3}}
\begin{abstract}
Vehicular ad-hoc networks, where traffic information is distributed from many sources to many destinations, require data authentication mechanisms to detect any malicious behavior of users, such as modification or replay attacks. In this paper we propose a new data aggregation protocol that uses probabilistic verification to detect such attack attempts a posteriori in an efficient way, with minimal overhead and delay.  The algorithm also contains an additional security mechanism based on reactive groups created on demand, which ensure a priori that vehicles generate trustworthy information. According to a comprehensive analysis including both a small-scale real device implementation and NS2 simulations, it is shown that the proposed protocol is robust.
\end{abstract}
\end{frontmatter}
\section{Introduction}
Vehicular Ad-hoc NETworks (VANETs) are usually defined as wireless networks formed among vehicles and roadside infrastructure, which are used to provide drivers with information to increase safety, efficiency and comfort in road travel. In this type of networks, warning messages affect decisions taken by drivers so that any false disseminated information could lead to higher transportation times, fuel consumption, environmental contamination and impact of road constructions, and, in the worst-case scenario, more traffic accidents.

As prospective intelligent transport system technology, VANETs have become a very hot topic in the research on networks \cite{FNSA12}. In the near future, this type of networks will allow the reduction of the number of deaths due to car accidents, and the provision of real-time information on traffic and on roads. For example, drivers will be able to exchange information with their neighbors and with the road so that they can receive warnings about potentially dangerous events such as accidents, obstacles on the road, etc. Another practical application of VANETs is, for instance, the ability to find free car parking spaces.

Nowadays, several GPS applications offer information of the traffic on a chosen route and compute alternative routes based on feedbacks from local road authorities, police departments and systems that track traffic flow. However, in most cases the information given to the driver is not real-time because it does not reflect the events that have just produced, and/or implies the lack of user privacy. Besides, most of that software, like Google traffic application, requires a 3G connection, what represents an additional cost for the users.
Thus, our motivation to study the secure, efficient and self-organized deployment of VANETs to assist drivers instead of those GPS applications is clear.

A classical VANET is composed of two different types of nodes:
On-Board Units (OBUs) that are wireless devices installed on
vehicles, and Road-Side Units (RSUs) that form the network
infrastructure on the road. While in other resource-constrained
wireless networks such as sensor networks, data aggregation is
used mainly to save energy \cite{Wang2011}, in VANETs data
aggregation can be used both to ensure that the transmitted
information is reliable and to minimize the number of repeated
event warnings \cite{MJ12}.

To overcome false content generation problems, a new data aggregation protocol is here proposed. In particular, we combine the ideas of reactive groups and data aggregation with a probabilistic verification scheme to check the authentication of warning messages quickly and reliably in self-organized and decentralized VANETs.

\subsection{State of the Art}

Regarding the protection of VANET communications, in the literature we can find several papers proposing the use of asymmetric cryptography in VANETs so that thanks to the use of digital signatures, both the source and the integrity of messages can be verified~\cite{Gollan}. Other works propose the use of symmetric encryption to provide location privacy~\cite{Wasef}. We can also find proposals based on the use of pseudonyms to protect user identities~\cite{Calandriello}. However, none of those mechanisms protect the system against malicious attacks such as false content packet generation, which is one of the objectives of the present paper. A legitimate but adversary node could try to inject false information that does not correspond to what it is really detecting. For example, a driver who wants to reach its destination as soon as possible might try to disseminate information about a false congestion on a road in its route in order to decrease the number of vehicles on it. To face this problem, the system described in~\cite{Daza} uses a mechanism based on threshold signatures, which prevents internal attackers from attempting to send fake messages. Three privacy-preserving variants of the system are there described to provide message trustworthiness and vehicle unlinkability under different road conditions. However, such a proposal requires the participation of a trusted governmental authority, which is not available in fully distributed and decentralized networks like the ones here analyzed. More recently, the paper \cite{PFGA12} proposes a method to hinder the dissemination of false warning events in VANETs by applying  cryptographic certificates.
Other works  propose data aggregation to address the same problem but under conditions from the ones of this paper. For instance,~\cite{OX2009} discusses the relationship between security and data aggregation in wireless sensor networks. 

The topic of this paper is about the need to act when false information is sent in VANETs. Thus, we are talking about trust management. Related to this topic, the survey~\cite{Jie2011} discusses the challenges, identifies some desired properties towards effective trust management and concludes the lack of effectiveness of the existing models. On the other hand, with respect to malicious attack identification, a work that discusses this topic is~\cite{Philippe2004}, where each node compares the received data with the stored information, as we assume here, but in that paper it is assumed that each vehicle has the global knowledge of the network, condition that is not assumed here.

With respect to data authentication in VANETs,~\cite{Zhang2008}
introduces a novel message authentication scheme that makes the
RSU responsible both for verifying the authenticity of messages
sent from vehicles, and for notifying the results back to the
vehicles. On the contrary, the model proposed here does not
require any RSU. Another proposal can be found
in~\cite{Eichler2006}, where the aggregation of multiple messages
describing the same event and the use of revocation messages
allowing vehicles to report false information are proposed.
However, such a mechanism has an important weakness because real
messages can be also revoked. In~\cite{Picconi2006} the proposed
solution is based on the use of a tamper-proof device and consists
in asking an aggregator vehicle about a random aggregated record.
The main disadvantage of this method is the dependency on a
tamper-proof device since an attacker could easily skip this
service in order to compose malicious aggregated data.
~\cite{RAH2006} proposes another mechanism to provide security
through aggregation in a scheme where streets are divided into
fixed size segments corresponding to Wi-Fi signal coverage. The
authors of~\cite{Wischhof} also outline an aggregation scheme that
combines all known information on each fixed-length road segment
to one average value. However, both aggregation criteria use a
fixed segmentation of the road, what has been shown that does not
work properly with a high number of vehicles in large areas, like
for example in big traffic jams covering kilometers.

Recently, the authors of~\cite{Dietzel2010} proposed the notion of data-centric trust for event validation, but their scheme produces a high dissemination delay. Another recent work closer to the present paper is~\cite{LSM10}, which proposes an algorithm to choose one of multiple aggregates for the same area based on a probabilistic approximation to underlying data. The difference with our proposal is clear because we propose a probabilistic approach applied on the verification of aggregated data, and not on the aggregation phase. Besides, they use aggregation to combine observations concerning large areas into one single value, instead of several aggregated packets produced by different reactive groups, which is one of the bases of the proposal here described.

Finally, regarding the formation of groups of vehicles, which is also a topic discussed in this work, there are many papers with different proposals. Thus, for instance,~\cite{Ibrahim2007} presents a protocol for relaying information under the assumption that vehicles form groups on a highway, there called clusters, and some details about speed and traffic information are exchanged within nodes in the same cluster. In their proposal, aggregated information contains also the relative positions of all cars to a cluster head and an average speed, what is not necessary in the scheme here described. Their proposal also reduces the amount of data transmitted about a cluster of cars, but it does not include any mechanism for merging aggregates.

\subsection{Our Contribution}
\label{back}
A natural approach to address the data authentication problem implies providing nodes of a mechanism to store and process received information, including data about type of warning, road where it was created, traffic direction, and source node that generated the packet, among others. When a warning message is received by a node, the device has basically two options: either to alert the driver of the danger even if the information is not true, or do not alert the driver and wait to be able to compare the received data in order to verify the contents of the packet accuracy, although this delay might cause an accident. The first option might affect the driver's decision and result in a waste of its time and/or money if the information is not true. Furthermore, it can increase distrust about other messages from the network. Thus, the recommended option in VANETs is the second one, even though it implies comparing received data with other information packets received before and related to the same event but provided by different vehicles, what could cause a considerable delay until receiving a sufficient number of packets with the same content from different sources. Apart from such a delay, this option requires that the vehicles have a large storage space as well as a fast mechanism to compare different records. Therefore, the implementation of such an option must take into account that the waiting time should be short enough to warn the driver in time to avoid the problem, and large enough to ensure that the content of the information is true.

\begin{figure}[htb]
 \centering
 \includegraphics[scale=0.50]{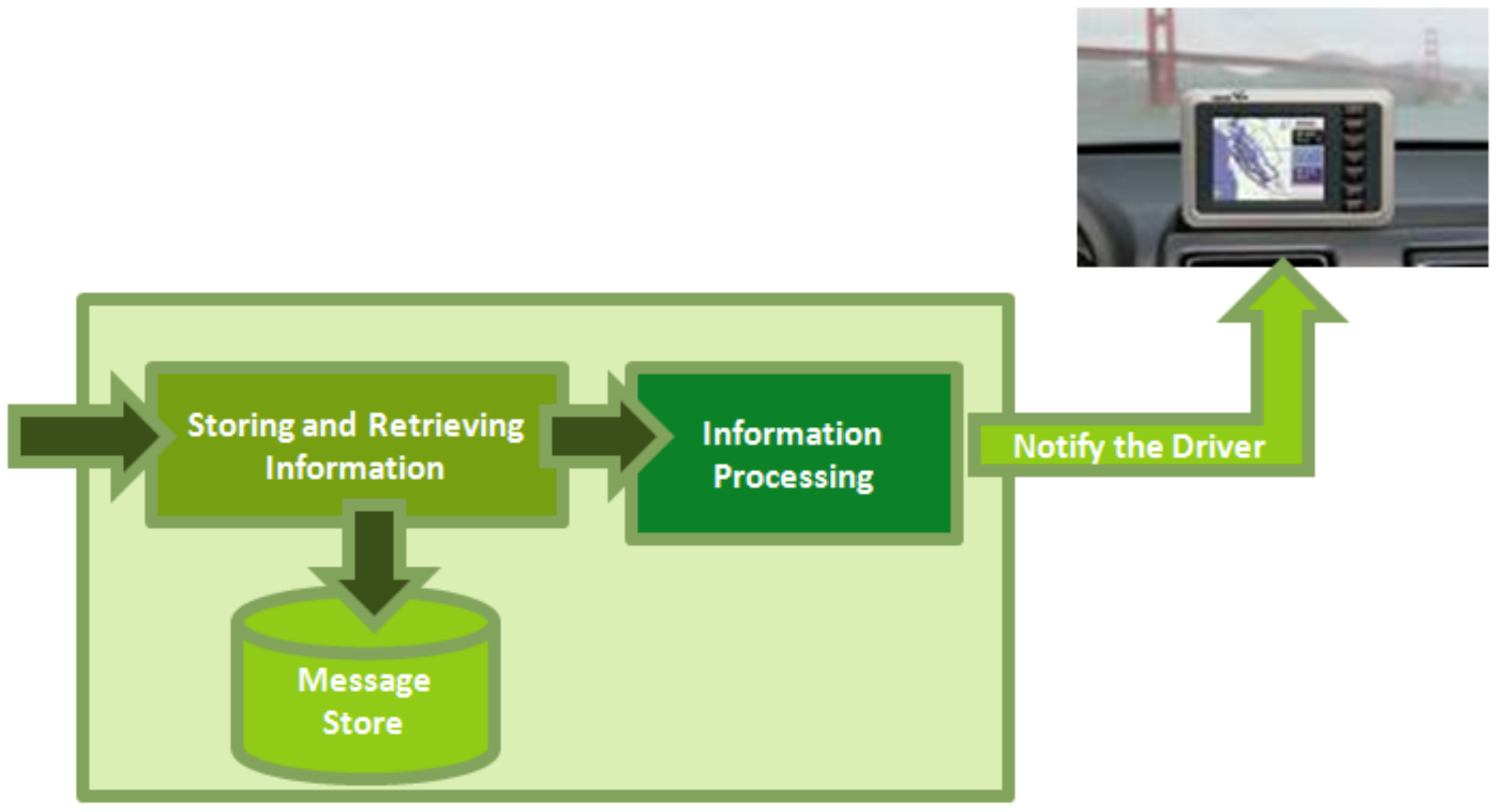} 
 \caption{Basic Model}
 \label{Fig:AgregacionB}
\end{figure}

In a basic model, each vehicle detecting an event, signs a warning message and broadcasts it, what means a considerable network overhead. Nodes that receive the signed packet, have to verify its signature and compare the content of the message with other related messages previously received, what also causes a major delay (see Fig.~\ref{Fig:AgregacionB}). To solve these problems, a simple model that combines the signatures generated by different vehicles to alert about the same danger can be used. However, the direct combination of signatures in a single packet would increase the packet size as the number of vehicles confirming the information increases. This would imply an overload of the channel too. Furthermore, the receiver in such a scheme must verify all the signatures, which also means another delay that would equal or even exceed the time required by the basic model.

In order to try to solve all the aforementioned problems a priori, we propose a new combined model based on two main ingredients. On the one hand, it uses reactive groups to provide trustful information through the limited combination of signatures in an aggregated packet. On the other hand, to solve the signature verification delay, we propose a probabilistic scheme that defines the verification of only a few signatures chosen at random among the signatures included in the packet. These two security mechanisms form the basis of the new model and will be fully detailed below.

This paper is organized as follows. In Section~\ref{sec2} the basic concept of geographic zones, details about reactive group formation and aggregated packet creation, and the probabilistic mechanism to verify data authenticity are introduced and discussed. The robustness of the proposed scheme is then confirmed in Section~\ref{ana} by considering both a theoretical analysis and different possible attacks. Section~\ref{sim} includes a performance evaluation through the analysis of results from real device implementation and  NS2 simulation. Finally, conclusions are presented in Section~\ref{conclusion}.

\section{Proposed Aggregation Scheme}
\label{sec2}
In this section we describe in detail a new data aggregation scheme that contains both a security mechanism based on reactive groups created on demand to ensure a priori that vehicles generate trustworthy information, and a probabilistic verification scheme to detect attack attempts a posteriori in an efficient way, with minimal overhead and delay.

\subsection{Geographic Zones}
\label{geo}

Now we introduce the definition of geographic zone, which is a key concept of the group-based mechanism of the proposal.

Due to specific characteristics of vehicular networks like high mobility and frequently changing topology, it is especially difficult to protect data in such networks. Thus, the security mechanisms in this environment should not assume the existence of any stable and centralized infrastructure, but only the existence of mobile functional nodes within the network. In our data aggregation scheme, we consider three different possible situations of nodes:

\begin{itemize}
\item Vehicles that find an event on the road and automatically generate a warning message.
\item Vehicles that receive a warning packet and can directly confirm that it corresponds to a true event.
\item Vehicles that receive packets with the event warning and its respective confirmation, but do not have direct contact with the reported event.
\end{itemize}

\begin{figure}[!h]
\centering
   {\epsfig{file = 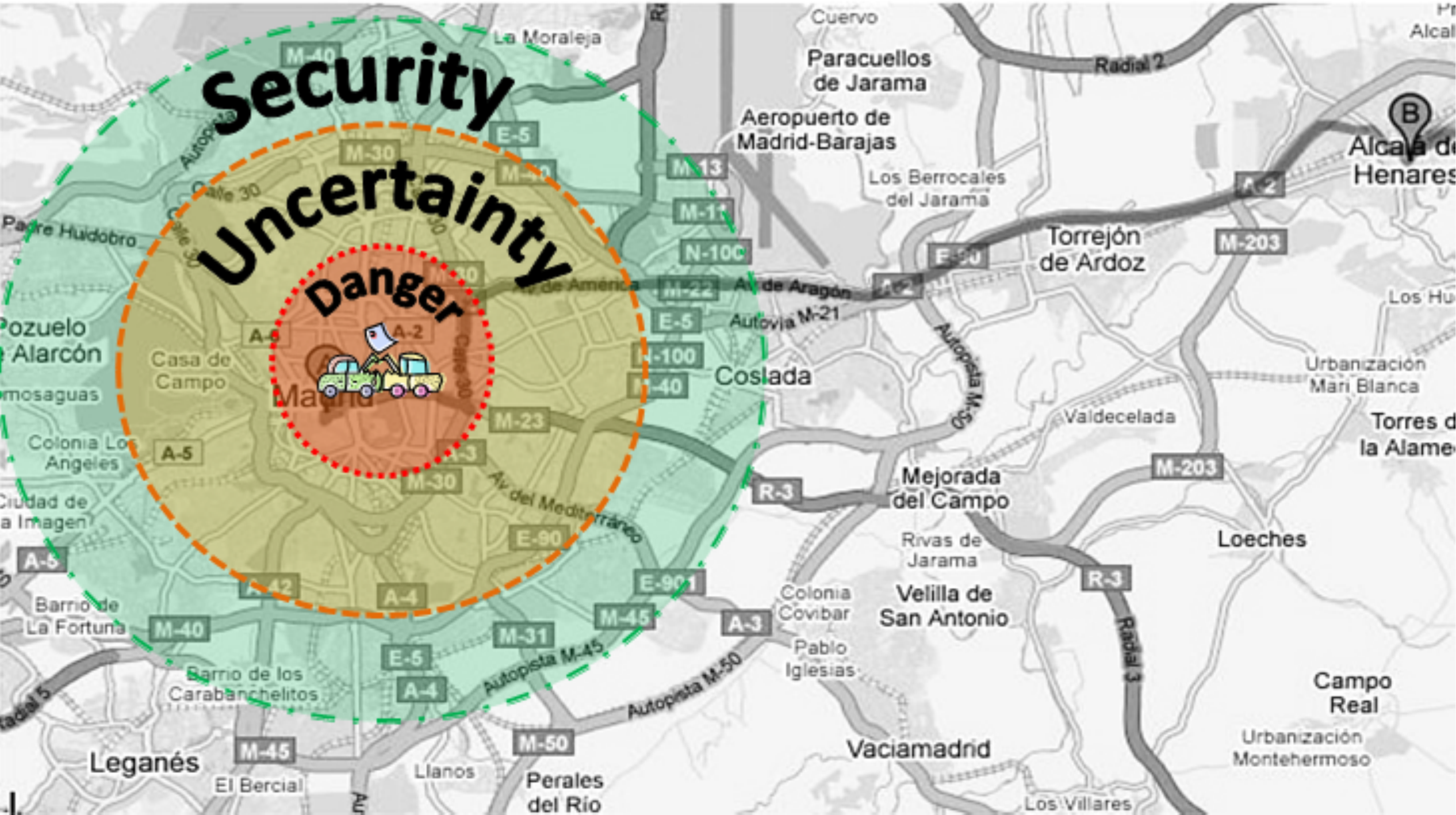, width = 9 cm}}
\caption{Geographic Zones}
\label{Fig:distance}
\end{figure}

In most cases, information generated at a certain location in a VANET is not interesting out of a radius distance. For example, if an accident happens in a city center, in most cases it makes no sense that the corresponding warning message reaches a neighbor city. Consequently, in this paper three different geographic zones are defined depending on where warning messages about an event is considered interesting by the receivers. In particular, different parts of the protocol must be run depending on the geographic zone where the receiver node is. As shown in Fig.~\ref{Fig:distance}, three geographic zones are defined with respect to a reported event:

\begin{itemize}
\item Danger zone, is the area defined by the innermost distance from the event, so that the event can be directly detected by vehicles in this zone.
\item Uncertainty zone, where nodes cannot confirm the information directly, but they have to make decisions quickly because in a short period of time they can be in the danger zone.
\item Security zone, where nodes cannot confirm the information directly, but they have enough time to collect evidences about the event in the form of aggregated packets.
\end{itemize}

The particular size of the radio of these zones is fixed by the source node, according to factors such as the type of road and event.

Aggregated and signed warning messages can be only produced by vehicles in the danger zone, while aggregated message verification is only run by vehicles that are unable to verify directly the information reaching them, that is to say, by vehicles out of the danger zone. In particular, when one of these vehicles receives a warning message about an incident that is not under the coverage of its antenna, and wants to confirm the authenticity of the received message, it has to act differently depending on the geographic zone where it is:

\begin{itemize}
    \item In the uncertainty zone, the decision must be made quickly to allow notifying the driver, who must be sure that the information is true. In this area, if a vehicle receives a signed aggregation packet, the vehicle should use a verification mechanism fast enough to verify all the signatures contained in the packet. However, as aforementioned, on the one hand it is inefficient to verify all the signatures contained in a packet, but on the other hand it is necessary to verify the information before accepting it as valid. In order to fix this problem, only a few signatures are proposed to be verified.

    \item In the security zone, since this region is quite far from the reported event, the receiving vehicle has more time to collect aggregated messages before it has to make a decision. In this case, the vehicle has to verify the signatures of the received packet, as in the previous case, but vehicles may also perform other verifications to provide a higher level of certainty on the received information. Being in this area, it is possible to receive different warning packets about the same event. Taking into account that these packets are independently generated, the more received and verified warning packets, the greater certainty and accuracy of the provided information. An additional verification that the vehicles in this zone can perform is to recreate the different cells where the received packets could be created from the data they contain, what provides value-added information because all the signatures in every packet should correspond to the same cell. This simple procedure gives an additional way to detect malicious nodes.

\end{itemize}

\subsection{Reactive Groups}
\label{groups}

In order to provide real-time and trustful information about events on the road, since no technical infrastructure is available to coordinate vehicles to act as a group, a reactive establishment of vehicle groups in a self-organized way is here presented. Thus, groups are formed only when they are necessary. This tool prevents any packet to grow indefinitely because it implies a limit in the number of signatures contained in it.
Here we propose a mechanism where group formation is not required a priori but when a vehicle detects an event, it automatically tries to form a group with other vehicles within its range in the geographic danger zone centered on the event. Note that otherwise, in dense environments, if all vehicles detecting an event sent the same warning message, the communication overhead in the network would be very high. In this way, the organization of vehicles in reactive groups to aggregate information allows avoiding repeated warnings. The event location defines the center of the danger zone, area that is proposed to be divided into cells to form groups. The leader of each group will be in charge of constructing the signed aggregated warning message.

\begin{figure}[!h]
\centering
   {\epsfig{file = 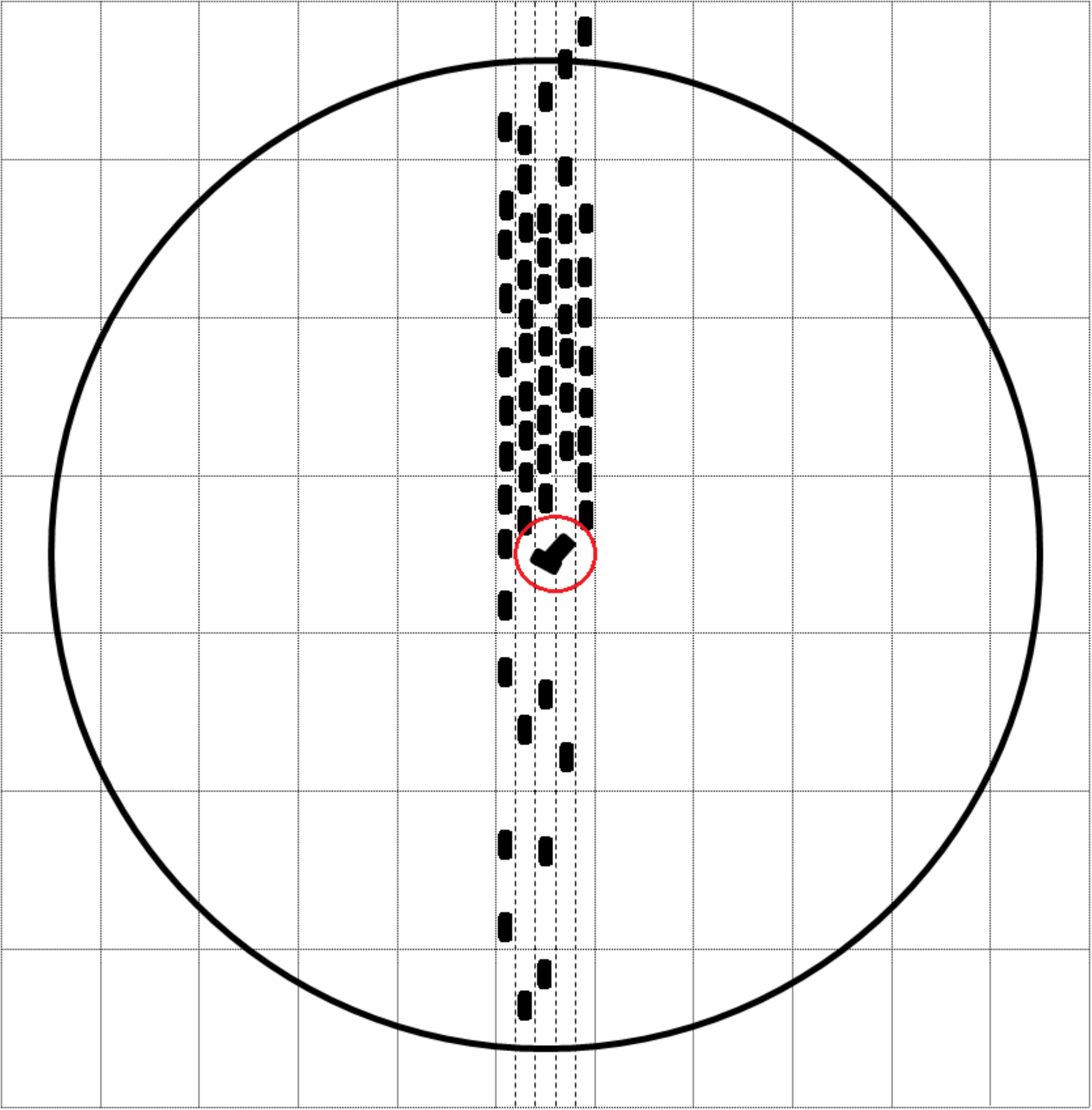, width = 8cm}}
 \caption{Cells in the Danger Zone}
 \label{dangerZone}
\end{figure}

As aforementioned, groups proposed in this work are reactive and created on demand because they are formed only once an event is detected on the road. In particular, when a vehicle detects a static event, it generates a packet with information about the event such as its geographic coordinates (X,Y,Z), timestamp, traffic direction, etc. This packet is broadcast to all nodes that are in the range of the danger zone. For instance, the radius could be 100 meters, which is roughly the transmission range of a Wi-Fi network. This danger zone is then divided into cells where reactive groups can be created as shown in Fig.~\ref{dangerZone}. Thus, a cell is a geographical area limited by the maximum number of vehicles that can form a group, while a group is a set of vehicles inside a cell, which produces an aggregated packet. In the maximum case, the number of vehicles in a group corresponds exactly to the maximum number of vehicles that fit into a cell. Since the dimensions both of the cells and of the danger zone are included in the message, all the receiver nodes in the danger zone define the same cells with respect to the event coordinates (X,Y,Z).

When a node receives a warning packet P from a node with ID, it runs the following process algorithmically described in Algorithm 1. First, it checks whether its vehicle is within the danger zone of the packet. Then, it checks whether it is able to detect the event, and in that case, it calculates the dimensions of the cell to which it belongs according to the coordinates (X,Y,Z) so that, if it has not received any group formation request, it launches a new one to all nodes within its cell. It is possible that two or more vehicles start the group formation request approximately at the same time, but this problem is easily solved as all vehicles always choose the request with the oldest timestamp and in the rare event of a timestamp collision, they choose the neighbor closest to the center of the cell. The chosen node becomes the leader of the reactive group and is in charge of generating the corresponding aggregated packet. Once the group is created, the leader signs the warning message and sends it to all members of its group.

When the vehicles of the group receive a signed warning message from the leader, they check whether they can validate the received information by verifying data such as time of the reported event and whether they are in the range to detect it. If they meet the above requirements but they do not detect the reported event, they might consider it as an attack attempt. In this case, the nodes would mark the leader node as malicious according to the scheme proposed in~\cite{{Mol2012}} and discard the packet. On the other hand, if they agree with the received information, they sign the message and send it back to the leader. The received signatures provide evidences so that the leader aggregates them to generate an aggregated warning message. This implies that the leader creates a packet where different vehicles from its group alert about the same event, what can be used as proof that the content of the information is true. Finally, the leader broadcasts this aggregated message and all the receivers verify the signatures according to the scheme described below and store the message.

\bigskip
\medskip
\smallskip
\textbf{Algorithm 1}  Reactive Group Formation
\hrule
\medskip

01: \textbf{function} Main(Packet P, Node ID)

02:\quad \textbf{bool} GroupFormationRequest= false;

03:\quad  ////Checks whether it is a new warning message

04:\quad \textbf{if} (Warning(P)) \textbf{then}

05:\quad \quad// Gets packet information

06:         \quad \quad  \textbf{double} X = P.X;

07:       \quad \quad   \textbf{double} Y = P.Y;

08:       \quad \quad   \textbf{double} Z = P.Z;

09:           \quad \quad  \textbf{string} EventType = P.EventType;

10:       \quad \quad \textbf{string} Direction = P.Direction;

11:       \quad \quad \textbf{string} Road = P.Road;

12:       \quad \quad   \textbf{double} TimeStamp = P.TimeStamp;

13: \quad  \quad  //Checks whether its vehicle is within the danger zone

14:     \quad  \quad \textbf{if} (InDanger(X,Y,Z,EventType,Direction,Road,TimeStamp)) \textbf{then}

15:     \quad  \quad \quad //Checks whether it is able to detect the event

16:     \quad  \quad \quad  \textbf{if} (DetectEvent(X,Y,Z,EventType,Direction,Road)) \textbf{then}

17: \quad  \quad \quad \quad  //Computes the dimensions of its cell

18:             \quad  \quad  \quad \quad \textbf{double} Cell = CellDimensions();

19:             \quad  \quad  \quad \quad // If it has not received any group formation request, from

20:                          \quad  \quad \quad  \quad \quad  \quad // its cell it launches one

21:             \quad  \quad \quad \quad \textbf{if} (!GroupFormationRequest()) \textbf{then}

22:             \quad  \quad \quad \quad \quad //Sends a group formation request

23:             \quad  \quad  \quad \quad \quad SendRequest(X,Y,Z,EventType,Direction,Road,TimeStamp);

24:             \quad  \quad  \quad \quad \textbf{end if}

25:         \quad  \quad \quad \textbf{else}

26:                 \quad  \quad  \quad \quad   MarkMalicious(ID); //Identifies attack attempt

27:         \quad  \quad \quad \textbf{end if}

28:     \quad \quad  \textbf{else}

29:         \quad \quad \quad //Cannot detect the event

30:     \quad \quad  \textbf{end if}

31:\quad \textbf{end Main}

\bigskip

It might happen that a node is alone in its group. In this case the node generates a packet with a single signature and sends it like an aggregated packet. As the packet has not enough evidence about the truth of the information, it is not taken into account in the uncertainty zone. However, this packet could be used combined with other warning messages about the same event received in the security zone.

\subsection{Types of Packets}

The described process of aggregation of data involves four types of packets:

\begin{itemize}
    \item Packet type W: warning packet broadcast by a source node who detects an event to all nodes in the danger zone. This packet contains the message M formed by (X,Y,Z) coordinates, type of event, traffic direction, name of road and timestamp, together with node ID and the signature of M using the private key $Pr$ of the node.

    \item Packet type R: corresponds to the request for group formation after the reception of a packet type W. This packet is sent by a node being proposed as leader of a reactive group to all the vehicles in its cell. In this case the content includes M together with the self-nominated leader's ID, signature of M, coordinates and timestamp in order to allow other nodes to determine whether the sender node is the leader of the group or not. The node sending this packet can know whether another node is in its own group or not thanks to its geographic coordinates, so it waits for a fixed time to receive all the signatures of the nodes belonging to its group.

    \item Packet type S: contains the signature of each node who agrees with the received packet type R, both with the event warning and with the group information. It contains the message M and the data corresponding to the sender.

    \item Packet type A: containing all the signatures received in packets type S by the leader of a group from its members in order to provide higher evidence about the existence of a reported event. This packet is broadcast by the leader through the network.
\end{itemize}

\begin{figure}[!h]
\centering
   {\epsfig{file = 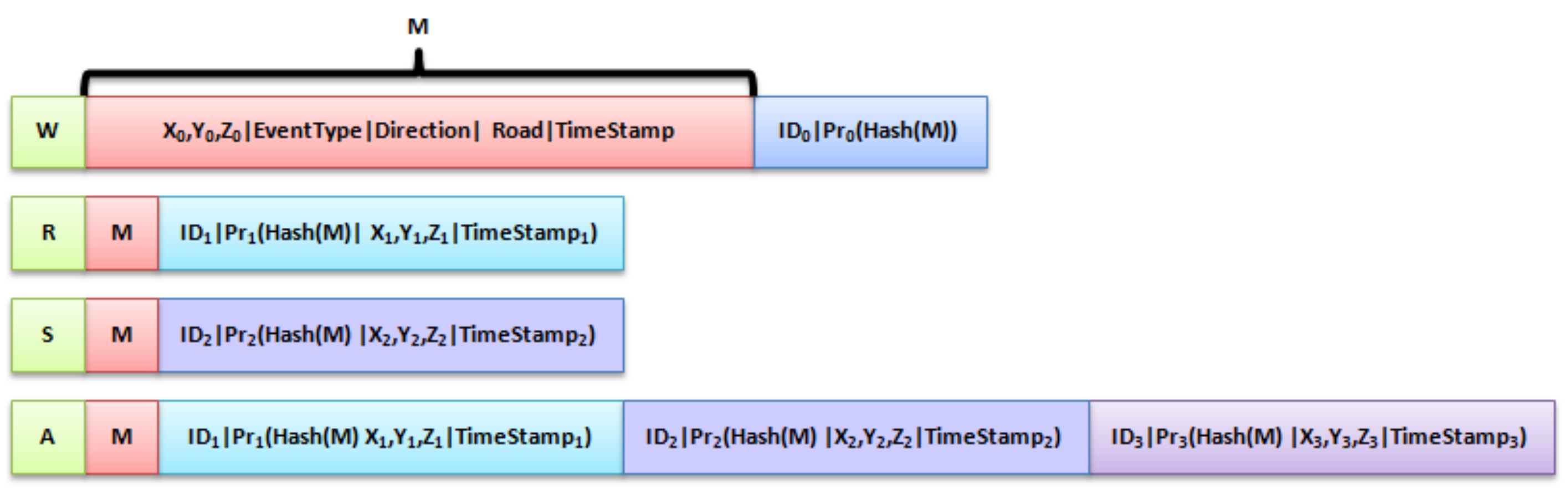, width = 13.5 cm}}
\caption{Types of Packets}
\label{Fig:packet}
\end{figure}

Fig. \ref{Fig:packet} shows a summary about the contents of each type of packet.

\subsection{Relaying Packets}

When a node receives a packet type A, it follows the store-and-carry paradigm~\cite{LW} and relays the packet each time it authenticates with another node. Thus, when two vehicles meet they follow an authentication protocol, and once it is completed both nodes exchange all  the events stored in their databases. The propagation of the messages is therefore in both directions. Note that on highways, the contact time between vehicles in different traffic directions is minimal so in this type of road, the largest packet exchange will be between vehicles traveling in the same direction.

In order to assure an optimum use of storage space, two decision parameters according to time and distance can be used to set limits on the storage of packet type A. This will prevent that events remain indefinitely stored in memory. The distance parameter was already mentioned in Section \ref{geo} where packets related to events outside their security zone are deleted. On the other hand, the use of a time parameter implies that the source node has to add to each event warning a timestamp indicating the time at which it detected it, as aforementioned. Since for instance a parking event is different from a traffic jam event, a different basic time $t_{E}$ must be considered for each type of event \textit{E}. Similarly, the type of road \textit{R} also influences the total time that has to be considered for event storage. Hence, for instance the factors $f_{R}$ for conventional road and for highway are different because the connection time in highway is generally shorter than in conventional road as the speed of vehicles is higher. Therefore, the storage time $t_{E,R}$ for each event \textit{E} on a road \textit{R} can be computed according to the formula:
  \begin{center} $t_{E,R} = t_{E} \cdot f_{R}$
\end{center}

For instance, in the simulations we used as basic time $t_{E} = 5$ min for traffic jam and $t_{E} = 90$ s for free parking space, and $f_{R} = 2$ for conventional road and $f_{R} = 1$  for highway. In any case, the optimal values of these parameters will be obtained after an extensive real device implementation.

\subsection{Cell Size}

The division of the danger zone into cells has been proposed here as a way to avoid channel overhead due to huge packets containing all the signatures of nodes detecting the same event. The aggregated packets produced in each cell should provide enough evidence about the validity of the reported event. Thus, the cell size must be chosen properly in order to address both issues because it is directly related to the packet size and to the number of signatures. In fact, it defines both the usual minimum and maximum number of vehicles inside it, what determines respectively the reliability and size of the corresponding packet type A.

According to~\cite{Ibrahim}, a fixed cell size of 16 meters wide and 126 meters long is the optimal cell size for VANETs. However, in a dynamic environment, these parameters should be variable in order to adapt them to the different environmental characteristics. In a VANET, vehicles are affected by different factors that vary while they circulate. Some of them are the number of lanes of the road, their speed, vehicle density supported by each road, etc.  Taking into account these changing characteristics, it is necessary to define specific criteria for determining the size of a cell based on this information.

It is impossible for every node to determine exactly how many vehicles exist on the road where it is at each moment, and the distance and speed of such vehicles, but each road has a maximum speed and minimum safety distance that vehicles must maintain, what gives an idea of the maximum number of vehicles that should be on it in normal conditions. Based on these parameters we can propose a metric to calculate the size of the cells.

The goal is to find an optimal cell size that minimizes the aggregated message size while maximizing the reliability of the information. In the simulations, we established that the length of each cell is twice the safety distance, value that can be easily calculated from the square of the road speed, while its width is the number of lanes of the road multiplied by the width of each lane, 4 m, what provides the area of the cell:

\begin{center} $Cell Area = (2 * safety\_distance)* (4 * number\_of\_lanes) $
\end{center}

This expression produces approximately the fixed optimal cell size indicated in~\cite{LW} for four-lane roads with 80 km/h speed limit.

In this way, the danger zone can be divided into cells so that the center of the danger zone (X,Y,Z) corresponds to the center of the central cell. The other cells are calculated from the central cell as shown in Fig.~\ref{dangerZone}.

The number of signatures that can be generated within a group in any of those cells is easy to compute. For instance, in a three-lane road with 120 km/h speed limit, CellArea= 3456 $m^{2}$ and a maximum of 9 signatures per cell are generated in the assumed conditions. Note that the influence of CellArea value when conditions are not those of the approximation formula is high. For instance, in the event of high density like traffic jam, approximately 170 signatures might be generated in the same cell, whilst in low traffic density, it might happen that only one vehicle exists in a cell and no more signatures can be generated there. However, if those traffic conditions are not stable, we must not either reduce the cell size to reduce the number of possible signatures in the cell because vehicles could move too fast to allow group formation, or increase the CellArea to increase the number of possible signatures because that could lead to too large packets. On the other hand, CellArea formula should not be dependent on the traffic conditions because such conditions may vary over time, and its value has to be clearly defined using only geographic coordinates.

A possible solution to the high density case, where too many signatures are generated in a cell, could be based on a probabilistic approach to data aggregation  in such a way that instead of generating a complete aggregated packet, only a few signatures chosen at random by the leader from the nodes in its cell are included in the probabilistically aggregated packet so that its size is decreased.

\subsection{Probabilistic Verification}
\label{prob}

The probabilistic verification of the signatures contained in a packet type A is the second basis of our proposal. In particular, the probabilistic aspect is on the choice of the signatures to be verified. Once the signatures are chosen, their verification is done through the usual method based on a Public Key Infrastructure (PKI), which involves obtaining the public key certificates of the signatures, and performing three steps: computation of the hash value of the signed message, decryption of the digital signature with the sender's public key, and comparison between both values. Since in VANETs it is not possible to rely on mechanisms that require a centralized system such as a centralized Certification Authority, public key certification must be done through some distributed solution like the one used in the implementation of~\cite{Patent2010}.

Probabilistic verification is only used by vehicles out of the danger zone, because those vehicles are unable to verify directly the information that reaches them and their only source of information in through the received packets type A. The proposed probabilistic verification algorithm uses threads, which are lightweight processes that allow a concurrent execution for a faster execution of the whole protocol.
In the algorithm shown below, Th[i] denotes a thread for the variable i that takes an integer value between 1 and $n$, where $n$ denotes the number of aggregated signatures in a received packet. When a vehicle receives an aggregated message, the main process launches as many threads as signatures the message contains but before that, it checks whether there are enough signatures to determine whether the message has been confirmed by a significant number of different vehicles. In the implementation, this minimum number of signatures was set to 3 but in general such a number should be fixed by the source node depending on the traffic density. In our particular simulation, if less than 3 signatures are received the packet is dropped, otherwise each thread Th[i] determines whether to verify the signature S corresponding to position i with a probability Prob(Th[i]), which is expressed using percent with a random number between 0 and 99. If Th[i] defines verification, and the signature is proved to be valid, Th[i] returns a true value informing that it is a valid signature. Otherwise, it returns a false value. The result of all these threads are stored in a structure St. If all fields in the structure St are proved to be valid, it is interpreted as evidence that all the verified signatures are correct so the message is accepted as valid.
On the other hand, if St contains some fields that are invalid, this would be interpreted as false message. Before signing a packet, legitimate nodes check whether they can validate the information. Otherwise, they discard the packet.

Prob(Th[i]) will be discussed in detail in the following section, where the reason  of its specific limit in the Algorithm 2 is given.

\bigskip
\medskip
\smallskip
\textbf{Algorithm 2}  Probabilistic Verification of Signatures
\hrule
\medskip

01: \textbf{function} Main(...)

02:\quad  bool St[n];

03:\quad  Thread Th[n];

04:\quad  \textbf{for} (i=0;i$<$n;i++)  \textbf{do}

05:\quad \quad ThreadStart[i].CheckSignature(n, St);

06:\quad  \textbf{end for}

07:\quad \textbf{if} (IsTrueMajority(St))  \textbf{then}

08:\quad \quad \textbf{return} ReliableMessage;

09:\quad \textbf{else}

10:\quad \quad \textbf{return} NotReliableMessage;

11:\quad \textbf{end if}

12: \textbf{end Main}

\bigskip
13: \textbf{bool function} CheckSignature(n, St)

14:\quad \quad \quad  \quad int j=0;

15:\quad  \textbf{if}(n $>$ 3)

16:\quad \quad \textbf{for} (i=0;i$<$n;i++)

17:\quad \quad \quad Prob(Th[i])=rand(0..99);

18:\quad \quad \quad \textbf{if} (Prob(Th[i]) $>$ ((1-10/n)*100)) \textbf{then}

19:\quad \quad \quad \quad string M=RecoverMessage(); //Gets the message

20:\quad \quad \quad \quad Signature S=RecoverSignature(i); //Gets a signature

21:\quad \quad \quad \quad St[j]=(VerifySignature(S, M));

22:\quad \quad \quad  \quad j++;

23:\quad \quad \quad \textbf{end if}

24:\quad \quad \textbf{end for}

25:\quad  \textbf{else}

26:\quad \quad //Not enough signatures to verify

27:\quad \quad \textbf{return} NotEnoughSignatures;

28:\quad \textbf{end if }

29: \textbf{end function}

\bigskip
30: \textbf{bool function} VerifySignature(Signature S, String M)

31:\quad  \textbf{if} (IsValid(S, M))  \textbf{then}

32:\quad \quad \textbf{return} true;

33:\quad  \textbf{else}

34:\quad \quad \textbf{return} false;

35:\quad \textbf{endif}

36: \textbf{end function}
\smallskip
\hrule
\medskip

\section{Analysis and Discussion}
\label{ana}
In this section an analytical evaluation of two important parameters of the proposal is described. In particular, the verification probability and the packet size are thoroughly studied. Also a brief discussion about possible attacks and countermeasures is included.

\subsection{Verification Probability}
To guarantee the validity of a specific message, a first approach would be that at least one thread should produce the verification of a signature. In this way, the probability that at least one thread leads to some signature verification must be as close to 1 as possible. However, from the aggregation's point of view, only one thread verifying a signature is not enough. Suppose that a message with several signatures is received so that only one of them is true and the rest are false. Then, if a thread Th[i] verifies only the true signature, it would ensure that this message is valid. Therefore, there should be more than one thread verifying the signatures of a message sent by a vehicle. In the following we analyze the optimal values to fulfill this restriction.

Each thread can be seen as an independent Bernoulli experiment that with probability $p$ given by the percentage Prob(Th[i]) produces the verification of the i-th signature of an aggregated packet~\cite{Feller}.
Let $X$ be the variable given by the number of successes of the $n$ threads which follows a binomial distribution with parameters $n$ and $p$. Thus, the probability of the event $B$ according to which there are at least two threads that verify the signatures of the packet can be expressed in function of $n$ and $p$:

\begin{center}
$P\left\{B\right\}=1-(1-p)^{n}-n\cdot p\cdot (1-p)^{n-1}$ \hspace{2cm}
(1)
\end{center}

Eq. (1) follows from that $P\left\{X=0\right\}=(1-p)^{n}$ is the probability that none of the signatures is verified and $P\left\{X=1\right\}=n\cdot p\cdot (1-p)^{n-1}$ is the probability that exactly one of the $n$ signatures is verified.

\begin{figure}[!h]
\centering
   {\epsfig{file = 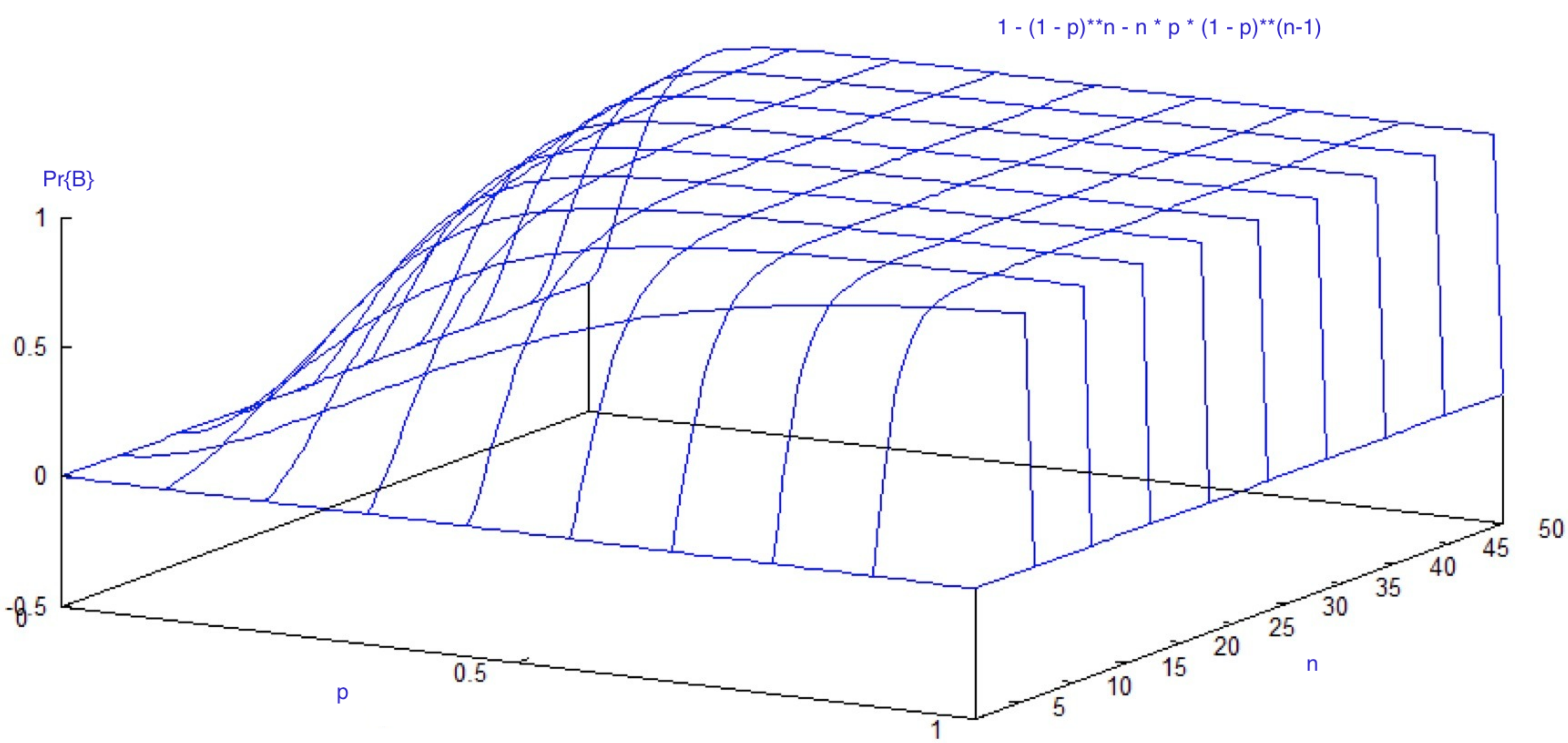, width = 12 cm}}
\caption{Probability of Verification of at Least two Signatures}
\label{Fig:graph}
\end{figure}

Our objective is to make $P\left\{B\right\}$ as close to 1 as possible. Fig.~\ref{Fig:graph} shows the relationship among $P\left\{B\right\}$, \textit{p} and \textit{n}. It can be seen that $P\left\{B\right\}$ increases as either \textit{p} or \textit{n} increases and quickly approaches 1. Our goal is to choose a \textit{p} value that makes $P\left\{B\right\}$ approaches 1 as much as possible for a fixed $n$ and at the same time is as small as possible because a small value of \textit{p} implies that a vehicle can potentially save processing time. In conclusion, the parameter \textit{p} must be adequately chosen so that both conditions are fulfilled. This balance is analyzed in the next section.

This work proposes an acceleration of the verification process through the addition of a probabilistic factor that implies certain risk that some false warning messages can pass the test because not all signatures are verified. However, since the parameters of the proposal are chosen so that at least two signatures are always verified, successful false warning messages can only be the result of a coalition of attackers. Since this work assumes that most nodes are honest, if the number of signatures in any aggregated packet is large enough, the probability of a successful attacker coalition is negligible.

\subsection{Packet Size}
\label{PS}
When the number $n$ of signatures contained in a packet is selected, both the maximum packet size that can be used for VANETs and the minimum number of signatures that is necessary for ensuring information must be taken into account. According to the first condition, packet sizes from 256 bytes to 1500 bytes might be considered in VANETs due to the capacity of the wireless channel. Since in such networks a large number of packets can be generated, it would be advisable not to use the maximum possible packet size because in that case a small number of packets can saturate the channel. In this paper we consider the use of about 100 bytes for the message content and the rest for the signatures, so that we can use for the signatures 156 bytes in the worst case and 1400 bytes in the best one. Given that the result of encrypting with private key almost does not change the size of the input, in this section we only take into account the result of applying the hash function to the message. For instance, the hash function SHA-1~\cite{SHA1} produces an output of 20 bytes, so that with it we could generate 7 signatures per packet of 256 bytes and 70 signatures per packets of 1500 bytes. These data are used as a starting point for the discussion on the optimal values of the parameters.

\begin{figure}[!h]
\centering
   {\epsfig{file = 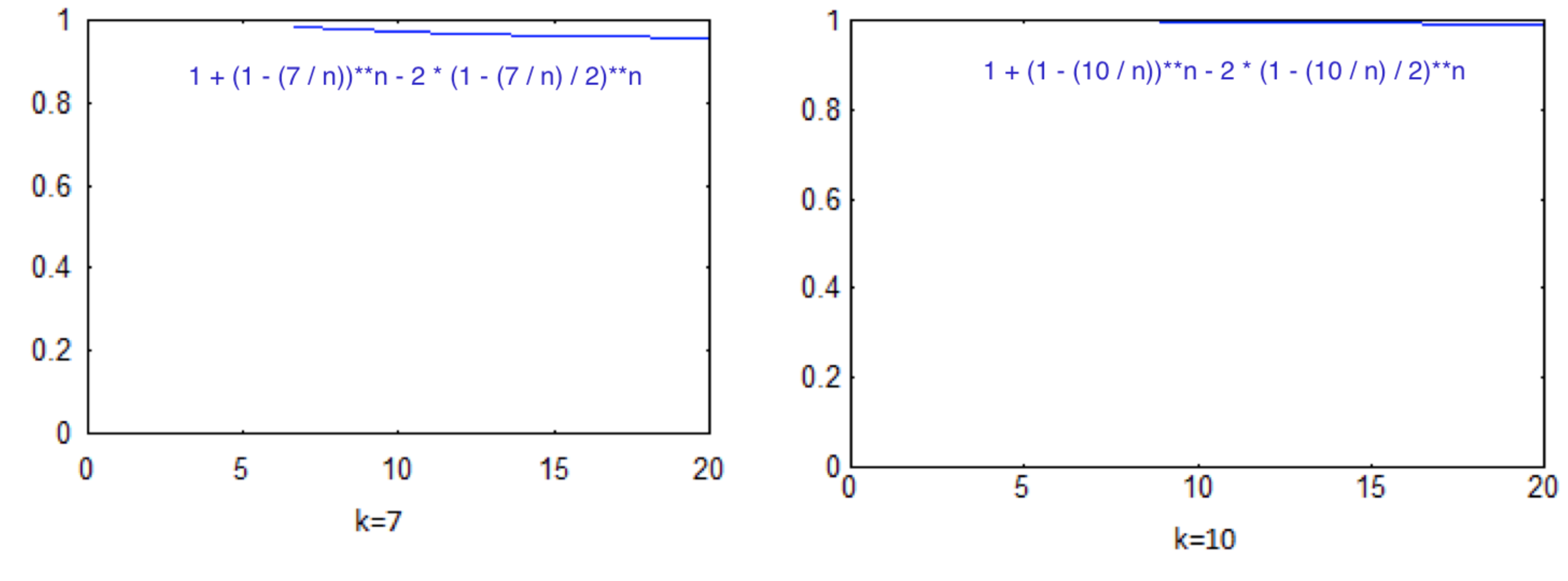, width = 12 cm}}
\caption{Probability of at Least 2 Verifications with $k$=6 and $k$=10}
\label{Fig:graphK7k10}
\end{figure}

In order to choose an appropriate value of \textit{p} for different values of \textit{n}, the variable $k=n\cdot p$ could be used to leverage the inversely proportional relationship between \textit{p} and \textit{n}. Notice that \textit{k} represents the average number of signatures that a vehicle verifies because \textit{n} is the total of signatures in the packet and \textit{p} is the verification probability. If we can find a suitable \textit{k}, then the corresponding \textit{p} can be determined. Based on Eq. (1), we can obtain the relationship between $P\left\{B\right\}$ and \textit{n} in terms of different values of \textit{k}, so that the value of \textit{p} can be determined. Given that the probability \textit{p} has a maximum value of 1, and we have that using SHA-1, \textit{n} would be greater than 6, we can use this value to conclude through Fig.~\ref{Fig:graphK7k10} that if we choose \textit{k=6}, $P\left\{B\right\}$ is close to 1 but not enough.  However, with k=10, $P\left\{B\right\}$ is sufficiently close to 1 when the packet contains 10 or more signatures. Therefore, we can conclude that $k$ can be set to a constant value, for instance $k=10$ and once $k$ is fixed, $p$ can be computed from $k/n$ so that it takes the value $10/n$, which is the limit used in the Algorithm 1. In other words, we can express $p$ in terms of the value of $n$. For example, a vehicle that receives a message with 20 signatures, verifies each signature with probability 0.5. Thus, if $n$ is less than 10, $p$ is $\approx$ 1, what looks correct because with so little evidence of an event, all signatures should be verified to avoid potential attacks.

Considering the minimum number of signatures that a packet can contain to maximize the probability $P\left\{B\right\}$, and calculating the probabilities and the maximum number of signatures that fit in a packet, we can discuss the optimal hash functions to be used in this type of network with the proposed scheme. Table 1 shows the corresponding data for different hash functions, MD5~\cite{MD5}  producing 16 bytes, SHA-1 producing 20 bytes and SHA-256 producing 32 bytes. Consequently, if 100 bytes are used for the message content, the table shows the maximum number of signatures that can be appended in each aggregated packet. For example, a packet of 512 bytes can include a total of 20 signatures with SHA-1. Even the use of the hash function SHA-256 with 512-byte packets is possible and increases the security of the scheme.

\begin{table}[ht]
\caption{Parameters of Hash Functions for a 100-byte Message}
\begin{center}
\begin{tabular}{| c | c | c | c | }
\hline \hline
\textbf{Hash} & \textbf{Packet} & \textbf{Signature} & \textbf{N. of } \\
\textbf{ Function} & \textbf{ Size} & \textbf{ Size} & \textbf{Signatures} \\
\hline \hline
MD5     &     &        & 9  \\
\cline{1-1}
\cline{4-4}
SHA-1   & 256 &     156 & 7\\
\cline{1-1}
\cline{4-4}
SHA-256 &     &         & 4\\
\cline{1-2}
\cline{3-4}
MD5     &      &         & 25  \\
\cline{1-1}
\cline{4-4}
SHA-1   & 512  &     412 & 20\\
\cline{1-1}
\cline{4-4}
SHA-256 &      &      & 12\\
\cline{1-2}
\cline{3-4}
MD5     &      &       & 57\\
\cline{1-1}
\cline{4-4}
SHA-1   & 1024 &     924 & 46\\
\cline{1-1}
\cline{4-4}
SHA-256 &      &         & 28\\
\cline{1-2}
\cline{3-4}
MD5     &      &         & 87\\
\cline{1-1}
\cline{4-4}
SHA-1   & 1500 &    1400 & 70\\
\cline{1-1}
\cline{4-4}
SHA-256 &     &        & 43\\
\hline\hline
\end{tabular}
\end{center}
\end{table}

\subsection{Security Analysis}
\label{attack}
In order to analyze the effectiveness and robustness of our proposal, we performed a security evaluation of the proposal. This section briefly analyzes several possible adversarial attacks and how our system can resist them.

In particular, eight attacks are briefly discussed below: Sybil attack, false information generation, aggregated message discarding, false aggregated message generation, impersonation attack, and attacks on privacy.
\begin{itemize}
\item \textbf{Sybil attack.} This type of attack occurs when a malicious node creates different false identities in the system in order to get more influence in the network. Our system only allows one identity per node using as unique identity a parameter such as its telephone number, so this attack is impossible.

\item \textbf{Generating false information.} An attacker may forge a message that does not correspond to the true characteristics of its real environment information. This case is dismissed by the data aggregation structure because the other vehicles sign the message only if they detect the same event and conditions that are specified in the message.

\item \textbf{Discarding aggregated messages.} Attackers may try to discard some aggregated message, resulting in biased information dissemination. To solve this problem some cooperation scheme could be used like the one proposed in~\cite{Mol2012}. Anyway, the damage that may result from the removal of one or a few data aggregation packets is not very high, since more than one aggregated packet related to each event are usually generated in the proposed system.

\item \textbf{Modifying aggregated messages.} An attacker might modify aggregated messages transmitted through the network. However, when a vehicle has no direct contact with the information contained in a received aggregated message because it is not in the danger zone, it has to perform some verification of the signatures. First, a vehicle in the uncertainty and security zones must verify that some signatures according to the verification probability match the message. Besides, the vehicles in the security zone must verify the existence of different aggregated packets from different reactive groups warning about the same event.
In any case, any false aggregated message would be detected with a high probability.

\item \textbf{Impersonation attack.} In order to protect the scheme against possible attacks consisting in claiming to be a legitimate node, robust public key management scheme and strong authentication are used in the implementation of the proposal.

\item \textbf{Attacks on privacy.} Privacy is an important concern in VANETs. Simple PKI-based communication systems do not protect nodes privacy because the broadcasting of any message usually includes the signer signature and certificate. Although these data do not contain any sensitive information on the sender, the receiver might be able to track it. This issue was addressed in the implementation of the proposal through the use of changing pseudonyms in beacon broadcast announcing presence so that to link a beacon to the corresponding nodes certificate is only possible after strong bidirectional authentication.  In this way, tracking all the movements of specific vehicles is not possible in the proposed scheme.

\item \textbf{False trust increase.} A legitimate node who does not detect an event could try to add its signature to the corresponding aggregated packet in order to increase the trust about the event. This attack is useless in the proposed scheme because the verification of only one signature is not enough to accept the validity of the warning message.

\item \textbf{Leader's attack.} A group leader could try to add a false signature to an aggregated packet about a true event in order to force the event to be discarded by the other nodes. Such an attack would be easily detected by the other nodes in the danger zone where they can correctly detect the event.

\item \textbf{Collusion attack.}  In this work, we assume that most nodes are honest. Thus, as aforementioned, in order to minimize the probability of successful coalition attacks, the number of signatures required to generate an aggregated packet must be large enough. Additionally, some cooperation scheme such as the proposed in \cite{Mol2012} could be used to prevent this type of attacks.

\end{itemize}

\section{Performance Evaluation}
\label{sim}
In this section an implementation analysis of the above proposal is included. To check the effectiveness of the aggregation and verification processes, the best option would be by testing them in a large scale implementation. However, doing this with a high number of real devices is not easy. Therefore, the chosen alternative has been to implement the scheme in a few devices to take real data from them and then use those data in NS2 simulations.

\subsection{ Real Device Implementation}
\label{impl}

We have implemented the proposal mechanism in a real VANET environment in order to obtain real data that are used in a software simulation. As aforementioned, the real device implementation of the proposal was combined with the implementation of public key management, strong authentication and pseudonym schemes in order to prevent several types of attacks.

VANETs are traditionally defined in bibliographic references as a set of special devices called On Board Units, which are installed in vehicles, and Road Site Units, which are deployed on the roads. These networks make possible the detection of events and the exchange of messages about such events between nodes. Both issues were assumed to be solved in 2011, but the economic crisis has prevented this VANET deployment because the aforementioned installations are too costly both for governments and for users.

In order to try to deploy VANETs without additional costs either on users or on governments, a new model has been proposed of a fully self-organized VANET where no RSUs are assumed and the OBU role is partially played by a software application called VAiPho (VANET in Phones)~\cite{Patent2010}, running on mobile phones in vehicles. Given that many references suggest that vehicle OBUs connect via Wi-Fi using 802.11 protocols, VAiPho implements VANETs in smartphones, because these are widespread devices that provide Wi-Fi connectivity, GPS and enough computing capability. In particular, VAiPho is a free secure communication system for spontaneous and self-managed vehicular networks that use smartphones and do not require any infrastructure in vehicles or roads. The operating mode is completely distributed and decentralized and takes into account the protection of privacy and integrity against possible attacks, as aforementioned.
Below the different modules implemented in VAiPho including the aggregation scheme here proposed are shown in Fig.~\ref{Fig:mod}:

\begin{figure}[!h]
\centering
   {\epsfig{file = 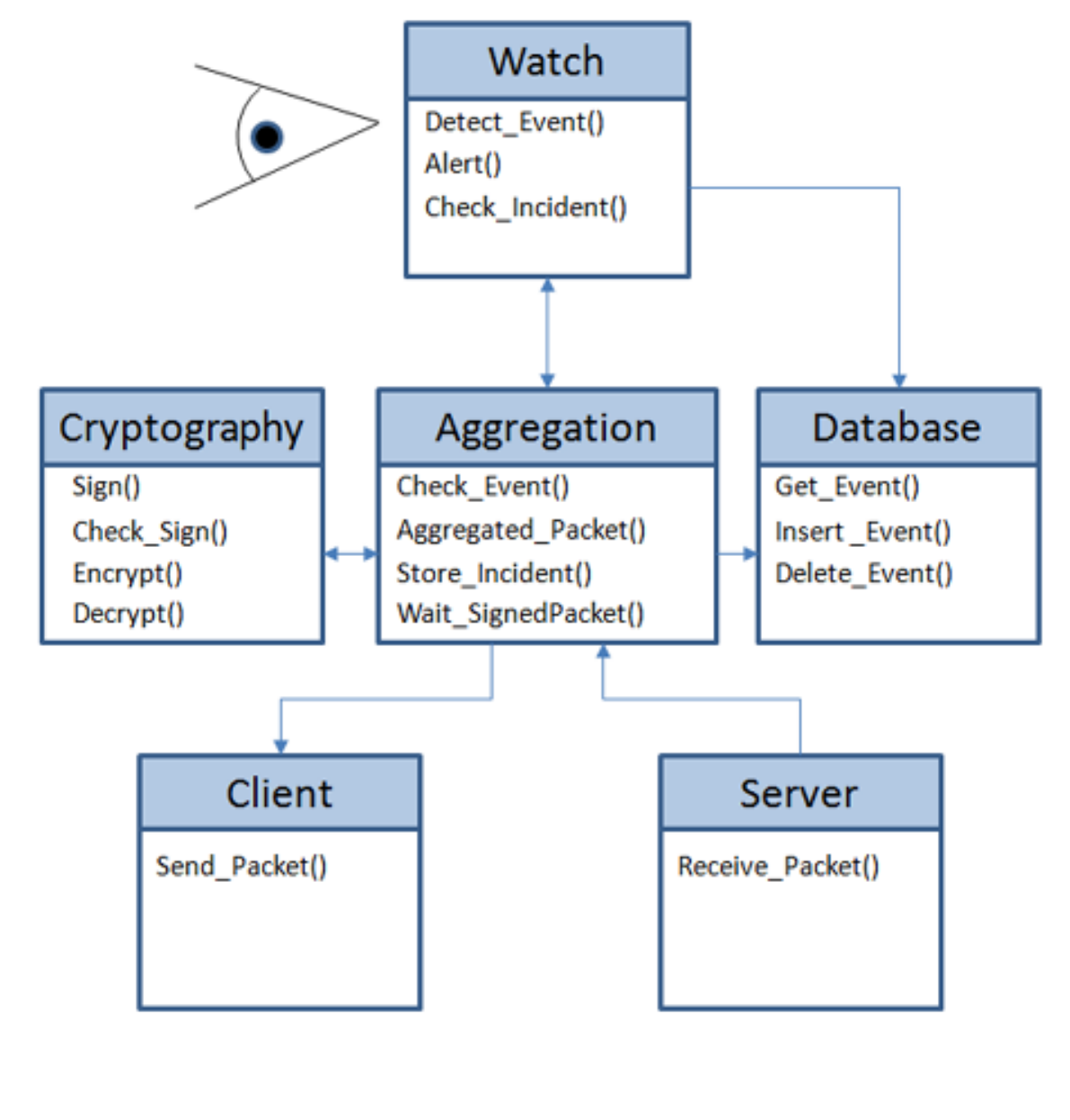, width = 11 cm}}
\caption{VAiPho Modules}
\label{Fig:mod}
\end{figure}

\begin{itemize}
    \item \textbf{Watch Module} is responsible for detecting events on the road, such as traffic jams. The module is continuously consulting the GPS in order to get the type of road, minimum and maximum allowed speed and car speed. If this speed is abnormally low according to the road speed, the device generates an alert about a possible traffic jam. After a period of time, if the vehicle continues at an abnormally low speed, it generates a traffic jam warning that launches the Aggregation Module and is stored in the database.

\item \textbf{Database Module} contains all the functions responsible for interacting with the database like event insertion, creation, consult and deletion. The database stores event warnings produced by the Watch Module or included in information packets received by the device.

\item \textbf{Aggregation Module} implements the aggregation functionality explained in this paper on an event warning produced by the Watch Module. The first task is to check whether the event indicated by the Watch Module exists or not in the database so that if it is not in the database, it means that it is a new event so it launches an aggregation procedure starting a group formation process.

\item \textbf{Client Module} sends messages either in broadcast mode or in unicast mode to a node, in order to form a reactive group with information about an event and waits until vehicles in the neighborhood sign the messages to form an aggregated packet.

\item \textbf{Server Module} is responsible for receiving and classifying packets into three groups: packets type R to be signed if the receiving node agrees with the reported event; packets type S with signatures to generate aggregated packets; and aggregated packets type A. Depending on the type of received packet, the module proceeds with the corresponding action.

\item \textbf{Cryptography Module} is responsible for signature, encryption and decryption of information.
\end{itemize}

\begin{figure}[!h]
\centering
   {\epsfig{file = 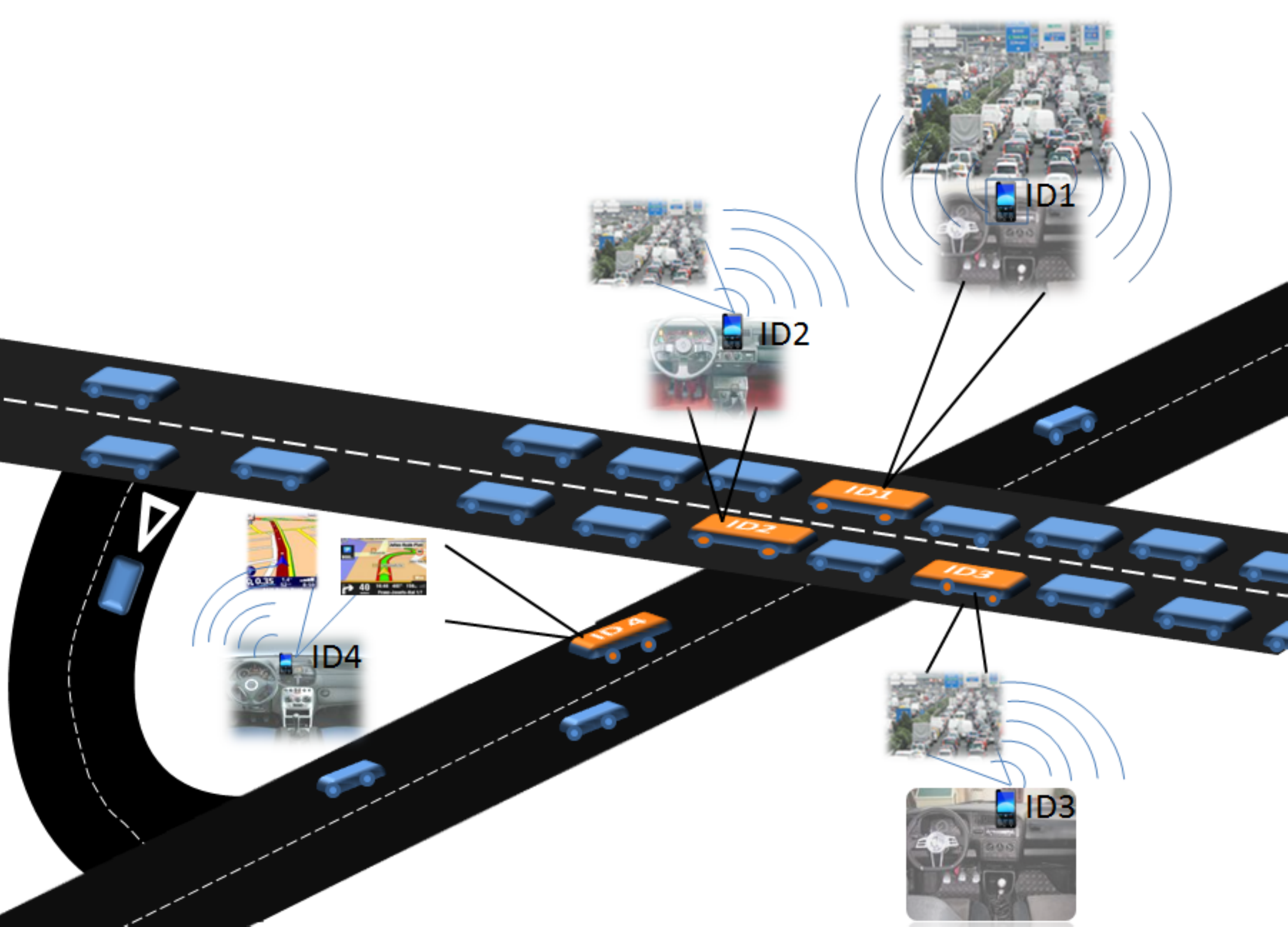, width = 10 cm}}
\caption{Simulated Situation}
\label{Fig:fig8}
\end{figure}

The real device implementation was made with four mobile devices each inside a different car, of which three were used to detect a traffic jam and create the corresponding signed aggregated packet type A after the corresponding exchange of the packets type W, R and S during the formation of the reactive group. The fourth device was inside a vehicle in the uncertainty zone so it could not see the reported event but received the aggregated packet type A and verified the signatures. In this case, since there were so few devices, they verified all the three signatures.  Fig.~\ref{Fig:fig8} shows in detail the simulated situation.

In Fig.~\ref{Fig:fig8} the three mobile devices capable of detecting the traffic jam are denoted ID1, ID2 and ID3. Node ID1 starts a data aggregation procedure by sending a packet type R to all nodes in the network through broadcast. Nodes ID2 and ID3 and even ID4 may receive this packet because they are in the neighborhood. However, only ID2 and ID3 are able to detect the traffic jam and belong to the corresponding reactive group, so they sign the warning message and send the corresponding packet type S back to node ID1. Once node ID1 receives the information, it generates a packet with the signatures of ID2 and ID3 and broadcasts the resulting aggregated packet type A.  When node ID4 receives the information, it verifies the signatures contained in the packet and if they are correct, it warns the driver and suggests an alternative route.

\begin{figure}[!h]
\centering
   {\epsfig{file = 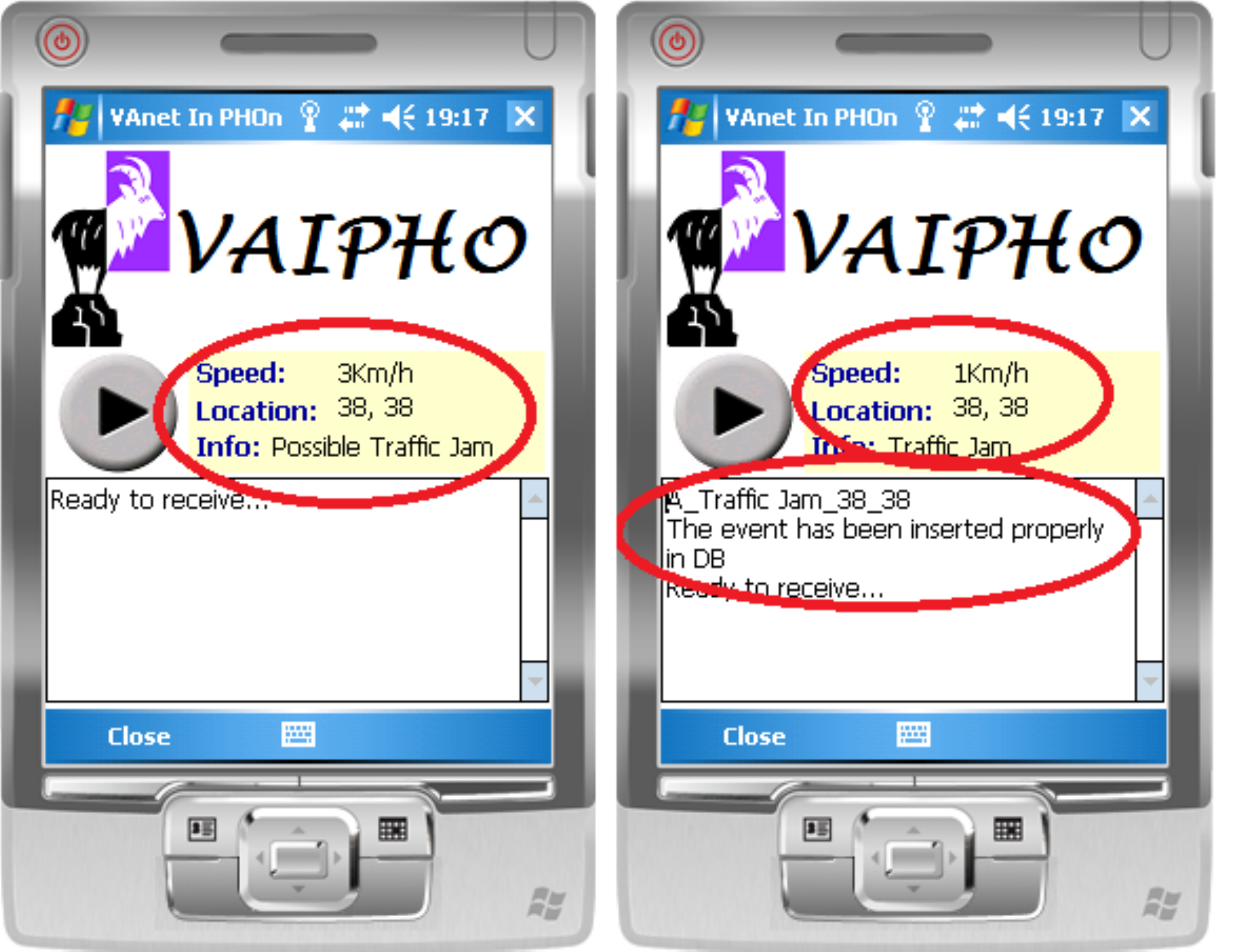, width = 10 cm}}
\caption{Traffic Jam Detection}
\label{Fig:Impl1}
\end{figure}

When the system starts, both the Server and the Watch Modules begin. As shown in the  image on the left of Fig.~\ref{Fig:Impl1}, the Watch Module collects the GPS information and when the vehicle speed is lower than 4 km/h, it indicates that there is a possible jam. After a few seconds it checks the conditions again in order to discard the possibility of traffic lights or a stop sign and if the situation remains equal, it inserts the reported event in the database as shown in the  image  on the  right of Fig.~\ref{Fig:Impl1}. All nodes are alerted but the process is only initiated by who first detects the event. In particular, the mobile node ID1 first detects a possible traffic jam and after certain time it checks again its speed to finally determine that it is a traffic jam. Then, the node stores the event in the database and sends a message with the information to form a reactive group. An aggregated packet is created with the signatures of all the nodes that form its reactive group and answer to its request.

\begin{figure}[!h]
\centering
   {\epsfig{file = 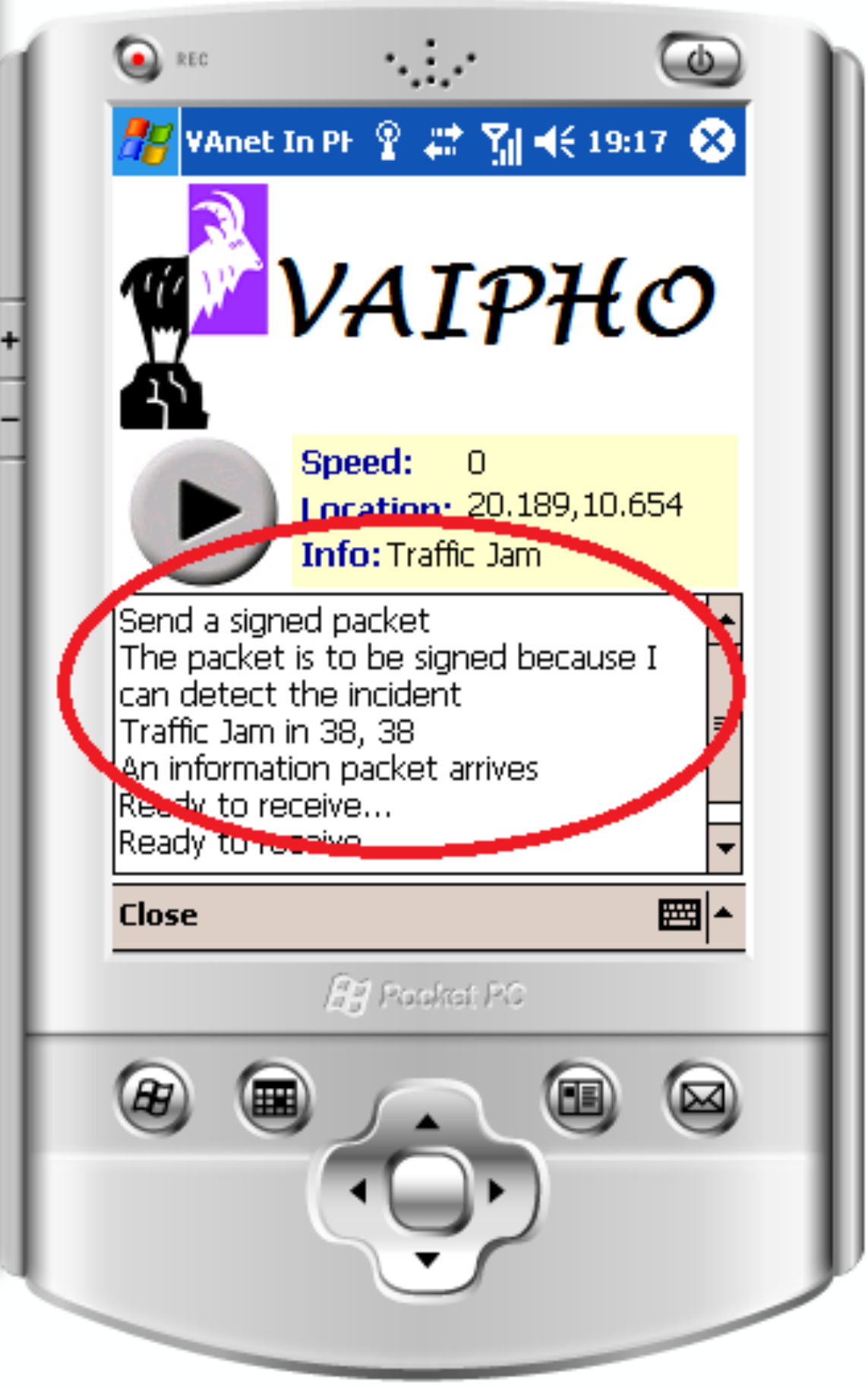, width = 5 cm}}
\caption{Traffic Jam Confirmation}
\label{Fig:Impl2}
\end{figure}

Nodes ID2 and ID3 receive the packet and check their Watch information. If they agree, they return a message with the signed information such as can be seen in Fig.~\ref{Fig:Impl2}. As explained above, the signature of the message is computed with the private key of each node and a hash function. Once signed, the message is forwarded to node ID1 indicating conformity with the received information.

ID1 receives two signatures from ID2 and ID3. Then, it generates the aggregated packet because there are no more devices in the reactive group. Finally, it broadcast the signed packet to the network so that it can be relayed to reach other vehicles that cannot detect the event directly.

\begin{figure}[!h]
\centering
   {\epsfig{file = 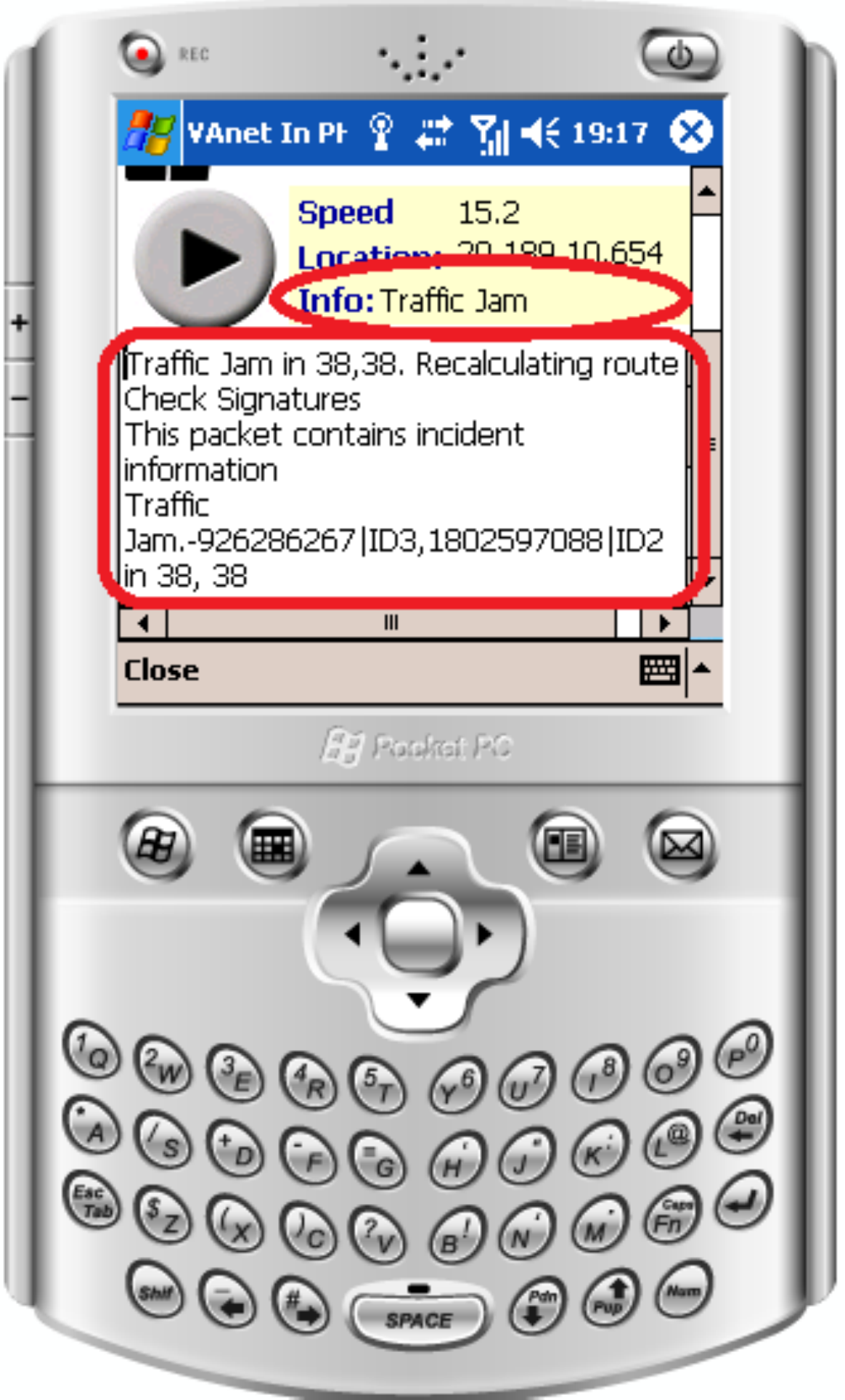, width = 5 cm}}
\caption{Signature Verification}
\label{Fig:Impl3}
\end{figure}

The aggregated packet is received by node ID4 who verifies the signatures before taking the information as valid. After such verification, the information is provided to the GPS to determine whether the vehicle should continue in the same route or take an alternative one. Fig.~\ref{Fig:Impl3} shows the coordinates of the reported event and how the warning is received by the device.
From the implementation we conclude that the proposed aggregation scheme works well in VAiPho solution for VANET deployment. Several videos explaining VAiPho operation and implementation details can be found on the website~\cite{Patent2010}.

Among the parameters obtained from the described real implementation, the most remarkable ones are the maximum generation time for packets type A, which was 2 min including the creation of the reactive group; the time required for the exchange of messages between mobile devices, which was less than 1 s; and the maximum distance between communicating devices, which was 300 m. The time for signature verification depends on the used hash function. All these data were used in the NS2 simulation described below.

\subsection{NS2 Simulation}
In order to analyze how fast and effective is the data aggregation module, several NS2 simulations have been done based on data got from the real device implementation. This section presents some details and results obtained by averaging 100 simulations using different network sizes over the same area of 1000 square meters, that is to say, considering different situations regarding traffic density. Due to computational constraints, our simulations are formed by networks between 10 and 40 nodes. The most relevant parameters selected for the demonstration have been: total number of lanes for each direction= 3, simulation time= 1000 s, moment when motions starts= 0 s, moment when retransmissions begins= 40 s, retransmission period= 10 s, transmission range= 100 m, traveled distance before the event happens= 800 m. Speeds and directions of nodes in the NS2 simulations were random. Two different types of roads and two different types of events were considered in such a way that the considered storage time of packets type A ranged from 3 min for a parking event announced in a conventional road,  to 5 and 10 min for a traffic jam event  in a highway and in a conventional road, respectively.

The aim is to evaluate on the one hand the number of generated packets using reactive groups, and on the other hand, the effects of our proposal in the computational complexity shown through the time that our aggregation mechanism  takes to warn all network nodes about the existence of an event. Finally, we analyzed the time spent in verifying the signatures contained in a packet for different packet sizes according to the proposed probabilistic verification. In the NS2 simulations, the used probability of verification of any signature is $p = 10/n$, such as it was already mentioned in Section~\ref{PS}. On the other hand, the number $n$ of signatures contained in each packet depends on the number of nodes that detect the event during the simulation. In particular, each packet contains at least three signatures and at most the maximum number of signatures that fits in each packet.

\begin{figure}[!h]
\centering
   {\epsfig{file = 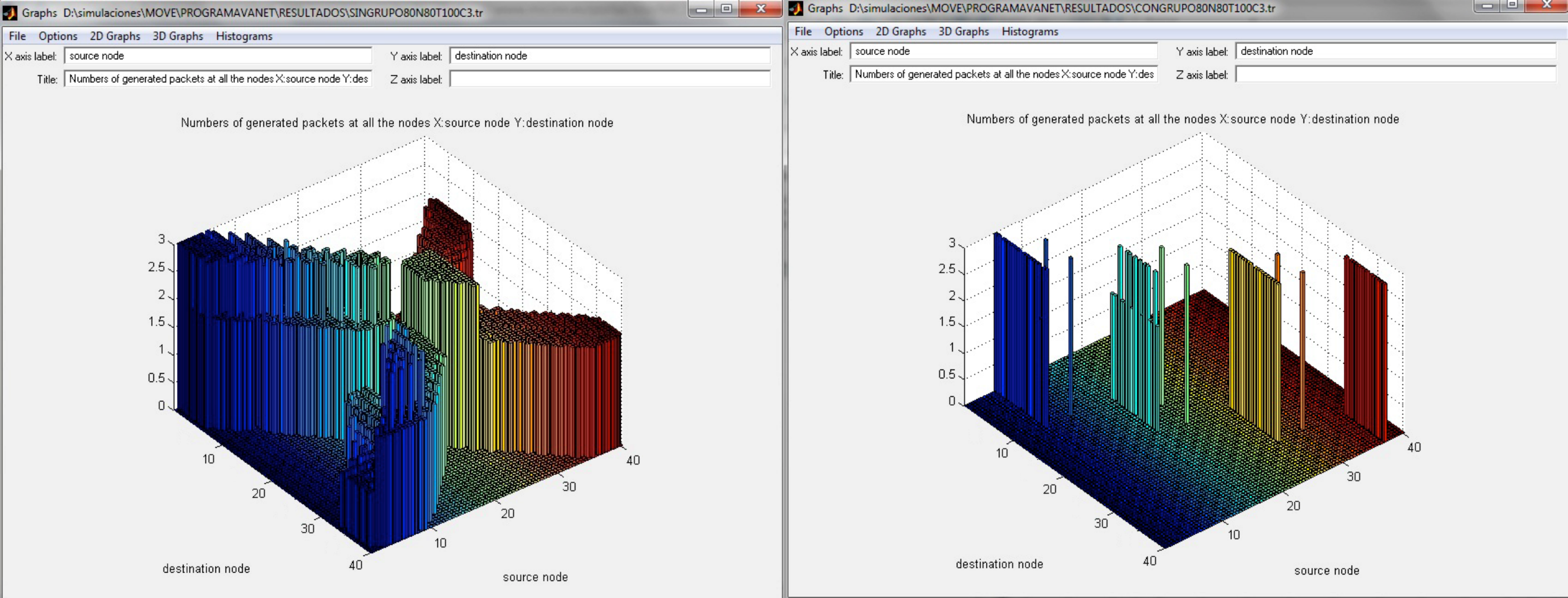, width = 13.5cm}}
\caption{Number of Generated Packets}
\label{Fig:fig6}
\end{figure}

The first simulations correspond to a traffic jam on the road and the corresponding warning packet forwarding with and without reactive groups, that is to say with and without aggregation. In Fig.~\ref{Fig:fig6} we can see that the number of generated packets in the simulation without aggregation is much higher than when using our aggregation scheme, even though in this case such a number includes the packets generated by the group formation algorithm. The decrease in the number of generated packets allows making better use of the channel. Thus, Fig.~\ref{Fig:fig6} shows how using aggregation with reactive groups we can reduce the number of generated packets and improve the channel usage.

Connections between vehicles in VANETs are usually quite short, so any proposed mechanism requiring communication between vehicles has to be fast enough to prevent data loss in communications. In Fig. 13 we can see the impact of node density in time cost of communication. For the purpose of these simulations, we compared the proposed aggregated scheme with a basic scheme without any aggregation mechanism. In such a basic scheme, each node that detects or receives an event, sends it to all nodes in its transmission range. We conclude that using this aggregation mechanism does not involve a significant increase in time cost of management when the packet is received. We can see that when the network size is between 10 and 20 nodes, the mechanism is even better. This effect is surprising but can be explained by the fact that communication is better organized when using aggregation and reactive groups, and permits to alert a greater number of nodes in less time. As the number of nodes in the network increases, the time it takes to process aggregated packet to warn all nodes also increases.

\begin{figure}[!h]
\centering
   {\epsfig{file = 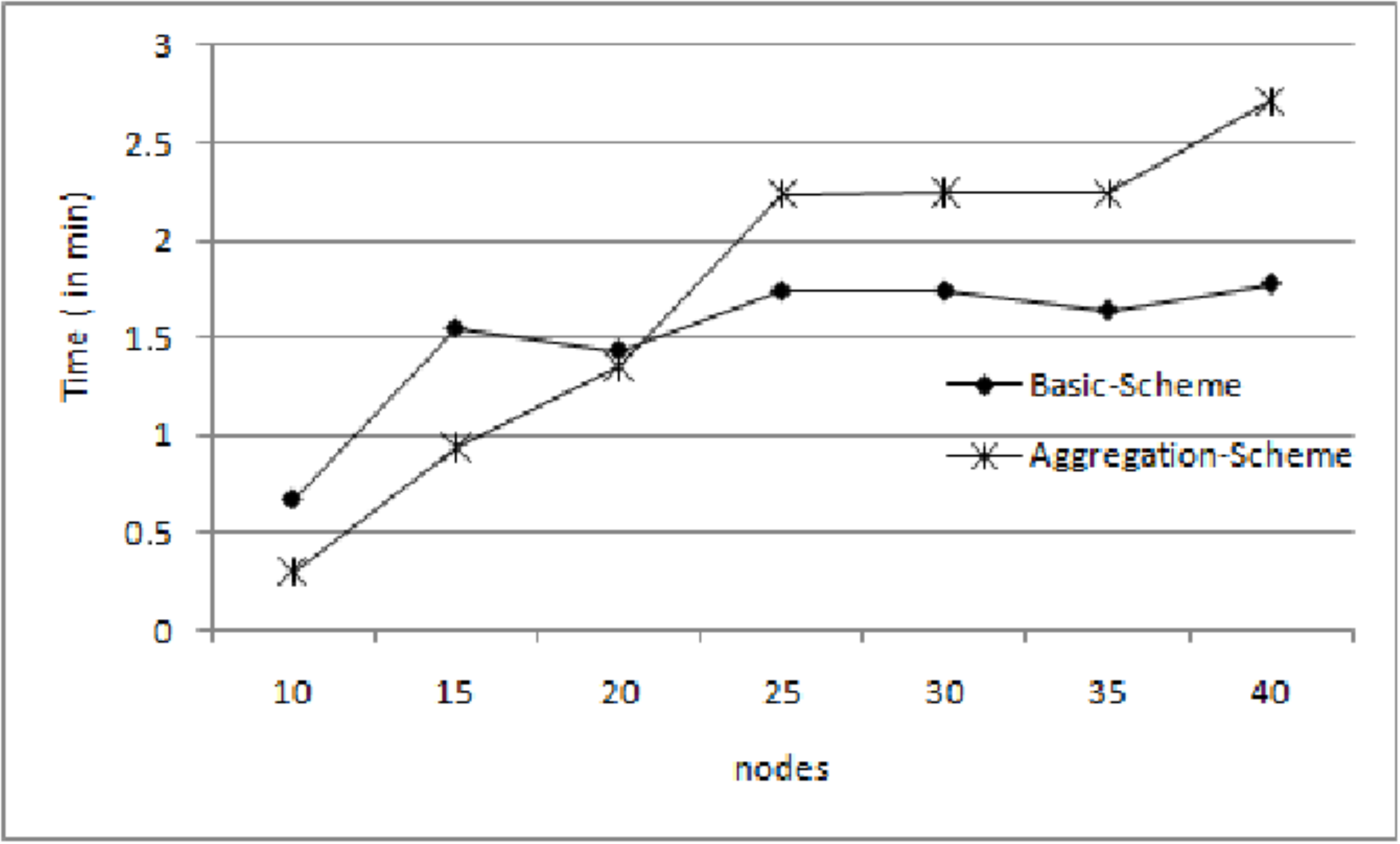, width = 12 cm}}
\caption{Time Cost}
\label{Fig:fig7}
\end{figure}

Another sensitive aspect of this implementation is the verification of signatures and the delay it causes. As we discussed in previous sections, it is impractical to verify all the signatures that a packet contains so in this work we propose a probabilistic algorithm in which only some of them are checked. To analyze the delay and find out whether our proposal improves the verification time we have made several simulations.
We took into account the different sizes of packets and the maximum number of signatures that fit in each one with the different hash functions mentioned in Section~\ref{ana}.

Table~\ref{Table:tab2} shows the simulation results. For different packet sizes and numbers of signatures, it shows the average times in minutes of 100 simulations. We can see that when the number of signatures contained in the packet is lower than 10, the test time is approximately equal for both methods although our method has a slightly higher time cost due to the calculation of probabilities. In the last column we can see the average of the number of signatures that our scheme verifies. Due to the proposed probabilistic approach, as the number of signatures increases, the time consumed by our scheme decreases, which makes our system more efficient. According to these results, our scheme maintains the verification time in about 0.138 min $\approx $ 8.3 s  regardless of the number of signatures contained in the packet. Therefore, we can conclude that the use of a probabilistic approach greatly improves the time spent on signature verification, and consequently demonstrates the scalability of the proposal.

\begin{table}[ht]
\caption{Simulation for Different Hash Functions }
\label{Table:tab2}
\begin{center}
\begin{tabular}{| c | c | c | c | }
\hline \hline
 & \textbf{Time} &  \textbf{Time } & \textbf{N. of }  \\
\textbf{Hash}  &  \textbf{Basic}&  \textbf{Aggr.} & \textbf{Verified}\\
\textbf{Function}  &  \textbf{Scheme}&  \textbf{Scheme} &  \textbf{Signatures}\\
\hline \hline
MD5     & 0.14121162    &   0.148480218 &   9 \\
\cline{1-4}
SHA-1   &  0.093453982 &        0.096037364 &7\\
\cline{1-4}
SHA-256  &0.058070086 &     0.064743644 &   4\\
\cline{1-4}
MD5      &0.379342788 &     0.133570748 &   10  \\
\cline{1-4}
SHA-1   &0.289680197 &      0.111626543 &   8\\
\cline{1-4}
SHA-256  & 0.182566924 &        0.171111552 &   10\\
\cline{1-4}
MD5      & 0.81221526 &     0.144320619  &9\\
\cline{1-4}
SHA-1  & 0.728451986     &  0.146179821 &   9
\\
\cline{1-4}
SHA-256 & 0.452038378 &     0.132944675 &   9\\
\cline{1-4}
MD5     & 1.246871085 &     0.135551543 &   11\\
\cline{1-4}
SHA-1   & 1.007615639 &     0.143795704 &   10\\
\cline{1-4}
SHA-256  & 0.679152238 &        0.135551543  &11\\

\hline \hline
\end{tabular}
\end{center}
\end{table}

\section{Conclusion and Future Works}
\label{conclusion}
This paper proposes a new solution to meet the need to address the security problem in VANETs consisting in determining whether road traffic information available to the driver is trustful or not. In particular, it describes a scheme to generate aggregated packets that cannot be replaced by any adversary because they contain the signatures of those vehicles who agree with the reported event. In order both to avoid that warning packets grow indefinitely because scalability is a primordial aspect in VANETs, and also to add trust to warning messages, signatures are generated according to reactive group formation. On the other hand, when an aggregated packet reaches a vehicle,  in order to avoid the delay produced by signature verification in dense environments, a probabilistic scheme is here proposed according to which only a few signatures are chosen to be checked.

In order to check the validity of the proposal, it has been implemented with real devices,  forming part of a software application for driving assistance called VAiPho, which has been developed to deploy VANETs with mobile phones inside vehicles.  The obtained feedbacks and results about VAiPho  have been overwhelmingly positive. In particular, the proposed system has been checked in a real environment where several devices share information about an event. These tests allowed us both to solve various real problems that do not appear in simulation environments, and to get data used in NS2 simulations, which also produced promising results regarding the time required by the scheme. It would be also interesting to validate our scheme not only with respect to the times to perform aggregation and verification, but also regarding its capacity to detect different types of attacks. Another future goal is the test of the accuracy of the proposed system in practice through a large scale real device implementation that allows us to  evaluate  the influence of speed and buildings in wireless communications.


\begin{thebibliography}{19}

\bibitem{Patent2010} Caballero-Gil, P., Caballero-Gil, C., Molina-Gil, J., 2010. VAiPho - VANET application for mobile Phones to avoid traffic jams, EUR, OTRI, University of La Laguna, Spain, Patent Number 201000865, http://www.vaipho.com.

\bibitem{Calandriello} Calandriello, G., Papadimitratos, P., Hubaux, J.P., Lioy, A., 2007. Efficient and robust pseudonymous authentication in VANET, Proc. of the 4th ACM International workshop on Vehicular ad hoc networks (VANETs), pp. 19-28.

\bibitem{Daza} Daza  V., Domingo-Ferrer, J., Sebe, F., Viejo, A., 2009. Trustworthy Privacy-Preserving Car-Generated Announcements in Vehicular Ad Hoc Networks, IEEE Transactions on Vehicular Technology, Vol. 58, No. 4, pp. 1876-1886.

\bibitem{Dietzel2010} Dietzel, S., Schoch, E., Konings, B., Weber, M., Kargl, F., 2010. Resilient secure aggregation for vehicular networks, IEEE Network, Vol. 24, Is. 1, pp. 26-31.

\bibitem{Eichler2006} Eichler, S., Merkle, C., Strassberger, M., 2006. Data aggregation system for distributing inter-vehicle warning messages, Proc. of the 31st IEEE Conf. on Local Computer Networks, IEEE Computer Society, pp. 543-544.

\bibitem{FNSA12} Faezipour, M., Nourani, M., Saeed, A., Addepalli, S., 2012. Progress and Challenges in Intelligent Vehicle Area Network, Communications of the ACM, Vol. 5, No. 2.

\bibitem{Feller} Feller, W., 1957. An Introduction to Probability Theory and Its Applications, Vol. 1, New York, John Wiley and Sons.

\bibitem{SHA1} FIPS 180-1, 1996. Secure hash standard, NIST, US Department of Commerce, Washington D.C., Springer-Verlag.

\bibitem{Gollan} Gollan, L., Meinel, C., 2002. Digital signatures for automobiles, Proc. of Systemics, Cybernetics and Informatics (SCI), pp. 225-230.

\bibitem{Philippe2004} Golle, P., Greene, D., Staddon, J., 2004. Detecting and correcting malicious data in VANETs, Proc. of the 1st ACM International workshop on Vehicular ad hoc networks (VANET), pp. 29-37.

\bibitem{Ibrahim2007} Ibrahim, K., Weigle, M.C., 2007. Accurate data aggregation for VANETs, Proc. of the 4th ACM International Workshop on Vehicular Ad Hoc Networks (VANET), pp. 71-72.

\bibitem{Ibrahim} Ibrahim, K., Weigle,  M.C., 2008. Optimizing CASCADE data aggregation for VANETs, Proc. of the IEEE International Workshop on Mobile Vehicular Networks (MoVeNet), pp. 724-729.

\bibitem{LW} Kolios, P., Friderikos, V., Papadaki, K., 2009. Ultra Low Energy Store-Carry and Forward Relaying Within the Cell, Proc. of the Vehicular Technology Conference Fall (VTC 2009-Fall), IEEE, pp. 1-5.

\bibitem{LSM10} Lochert C., Scheuermann, B., Mauve,  M., 2010. A probabilistic method for cooperative hierarchical aggregation of data in VANETs,  Ad Hoc Networks, Elsevier, Vol. 8, Is. 5, pp. 518-530.

\bibitem{MJ12} Mohanty, S.,  Jena, D., 2012. Secure Data Aggregation in Vehicular-Adhoc Networks: A Survey, Procedia Technology, Vol. 6, pp. 922-929.

\bibitem{Mol2012} Molina-Gil, J., Caballero-Gil, P., Caballero-Gil, C., 2012. Countermeasures to Prevent Misbehaviour in
VANETs,  Journal of Universal Computer Science, Vol. 18, No. 6, pp. 857-873.

\bibitem{OX2009} Olteanu, A., Xiao, Y., 2010. Security Overhead and Performance for Aggregation with Fragment Retransmission (AFR) Very High-Speed Wireless 802.11 LANs, Proc. of IEEE Transactions on Wireless Communications, Vol. 9, No. 1, pp. 218-226.

\bibitem{PFGA12} Palomar, E., de Fuentes, J.M., Gonz\'alez-Tablas, A.I., Alcaide, A., 2012. Hindering false event dissemination in VANETs with proof-of-work mechanisms, Transportation Research Part C: Emerging Technologies, Vol. 23, pp. 85-97.

\bibitem{Picconi2006} Picconi, F., Ravi, N., Gruteser, M., Iftode, L., 2006.  Probabilistic Validation of Aggregated Data in Vehicular Ad-Hoc Networks, Proc. of the 3rd ACM International Workshop Vehicular Ad Hoc Networks (VANET), pp. 76-85.

\bibitem{RAH2006} Raya, M., Aziz, A., Hubaux, J.-P., 2006. Efficient secure aggregation in VANETs, International ACM Conference on Mobile Computing and Networking, pp. 67-75.

\bibitem{MD5} Rivest, R.L., 1992. The MD5 message-digest algorithm, Request for Comments (RFC) 1321, Internet Activities Board, Internet Privacy Task Force.

\bibitem{Wang2011} Wang, L., Wang, L., Pan, Y., Zhang, Z., Yang, Y., 2011. Discrete logarithm based additively homomorphic encryption and secure data aggregation, Information Sciences, Vol. 181 (16), pp. 3308-3322.

\bibitem{Wasef} Wasef, A., Shen, X., 2010. REP: Location Privacy for VANETs Using Random Encryption Periods, ACM Mobile Networks and Applications (MONET), 2010, Vol. 15, No. 1, pp. 172-185.

\bibitem{Wischhof} Wischhof, L., Ebner, A., Rohling, H., 2005. Information dissemination in self-organizing intervehicle networks, IEEE Transactions on Intelligent Transportation Systems, Vol. 6 (1), pp. 90-101.

\bibitem{Zhang2008} Zhang, C., Lin, X., Lu, R., Ho, P.-H., Shen, X., 2008. An Efficient Message Authentication Scheme for Vehicular Communications, IEEE Transactions on Vehicular Technology, Vol. 57, No. 6, pp. 3357-3368.

\bibitem{Jie2011} Zhang, J., 2011. A Survey on Trust Management for VANETs, Proc. of the 25th IEEE International Conference on Advanced Information Networking and Applications (AINA), IEEE Computer Society, pp. 105-112.

\end{thebibliography}
\end{document}